\documentclass{aa}  

\usepackage{graphicx}
\usepackage[varg]{txfonts}
\usepackage{lscape}
\usepackage{amsmath}
\begin{document}

   \title{Deep SDSS Optical Spectroscopy of Distant Halo Stars}

   \subtitle{III. Chemical analysis of extremely metal-poor stars.}

   \author{E. Fern\'andez-Alvar\inst{1}
          \and
          C. Allende Prieto\inst{2,3}
      \and
          T. C. Beers\inst{4}
      \and
          Y. S. Lee\inst{5}
      \and 
          T. Masseron\inst{6}
      \and
          D. P. Schneider\inst{7}
          }

   \institute{Instituto de Astronom\'{\i}a, Universidad Nacional Aut\'onoma de Mexico, AP 70-264, 04510 Ciudad de M\'exico, M\'exico\\
             \email{emma@astro.unam.mx}
   \and
              Instituto de Astrof\'{\i}sica de Canarias,
              V\'{\i}a L\'actea, 38205 La Laguna, Tenerife, Spain\\              
         \and
             Universidad de La Laguna, Departamento de Astrof\'{\i}sica, 
             38206 La Laguna, Tenerife, Spain \\     
	 \and
		Department of Physics and JINA Center for the Evolution of the Elements, University of Notre Dame, Notre Dame, IN, 46556, USA \\  
         \and
         Department  of  Astronomy  and  Space  Science,  Chungnam  National University, 99 Daehak-ro, Daejeon 34134, Republic of Korea \\
     \and
         Institute of Astronomy, University of Cambridge, Madingley Road, CB3 0HA, Cambridge, United Kingdom \\
     \and
             Department of Astronomy and Astrophysics, The Pennsylvania State University, 
             University Park, PA 16802, USA \\
                          }

   \date{Received May 2016; accepted xxxx}

 
  \abstract
   {}
   {We present the results of an analysis for 107 extremely
metal-poor (EMP) stars with metallicities less than [Fe/H] $= -3.0$,
identified from medium-resolution spectra in the Sloan Digital Sky
Survey (SDSS). Our analysis provides estimates of the stellar effective
temperatures and surface gravities, as well as iron, calcium, and
magnesium abundances.}
   {We follow the same methodology as in previous papers of this series, based
on comparisons of the observed spectra with synthetic spectra. The abundances of
Fe, Ca, and Mg are determined by fitting spectral regions dominated by lines
of each element. In addition, we present a technique to determine upper
limits for elements whose features are not detected in a given spectrum.
We also analyse our sample with the SEGUE Stellar Parameter
Pipeline,  in order to obtain additional determinations of
the atmospheric parameters, iron and alpha-element abundances, to compare with ours,
and to infer [C/Fe] ratios.}
   {We find that, in these moderate to low signal-to-noise and medium-resolution spectra in this metallicity regime, Ca is usually the only
element that exhibits lines that are sufficiently strong to reliably measure its
abundance. Fe and Mg exhibit weaker features that, in most cases, only
provide upper limits. We measure [Ca/Fe] and [Mg/Fe] for EMP stars in
the SDSS spectra, and conclude that most of the stars exhibit the usual
level of enhancement for $\alpha$-elements, $\sim$+0.4, although a
number of stars for which only [Fe/H] upper limits could be estimated
point to higher [$\alpha$/Fe] ratios. We also find that 26\% of the stars in our sample can be classified
  as carbon-enhanced metal-poor (CEMP) stars, and that the
  frequency of CEMP stars also increases with decreasing metallicity, as has been reported for  
  previous samples. 
We identity a rare, bright ($g = 11.90$) EMP star, SDSS J134144.61$+$474128.6, with
[Fe/H] $= -3.27$, [C/Fe] = $+ 0.95$, and elevated magnesium ([Mg/Fe] $= +0.62$),  
an abundance pattern typical of CEMP-no stars.}
   {} 


   \maketitle
%

\section{Introduction}
\label{introduction}

Very metal-poor (VMP; [Fe/H] $< -2.0$) and extremely metal-poor (EMP;
[Fe/H] $< -3.0$) stars provide the opportunity to deepen our
understanding of the early chemical evolution of the Milky Way and the
Universe. In most cases, their atmospheres exhibit the chemical
compositions of the gas from which they formed, enriched by the
nucleosynthetic yields of the first stellar populations. In addition,
recent investigations of the Milky Way's stellar halo have revealed
that it is not a homogeneous entity, but rather,
comprises multiple populations (Carollo et al. 2007,
2010; de Jong et al. 2010; Nissen \& Schuster 2010, 2011; Beers et al.
2012; An et al. 2013, 2015; Allende Prieto et al. 2014; Chen et al.
2014; Janesh et al. 2016). Thus, VMP and EMP stars are useful probes of
the assembly of the Galaxy as well.

Early survey work, e.g., the HK survey of Beers and colleagues (Beers et
al. 1985, 1992) and the Hamburg/ESO survey of Christlieb and
collaborators (Reimers \& Wisotzki 1997; Christlieb 2003), provided the first large lists of
several thousand VMP and EMP stars. More recently, advances in
astronomical instrumentation has enabled multiplexed spectroscopy of
even larger numbers of (generally fainter) stars, e.g., the Sloan
Digital Sky Survey (SDSS; York et al. 2000), producing samples of VMP
and EMP stars covering a large range of distances from the Sun. 

High-resolution spectroscopic follow-up of these targets has provided
abundance measurements for numerous elements, revealing the existence of
chemical peculiarities that have greatly expanded our knowledge
of the different nucleosynthetic pathways that contributed to the early
chemical evolution of the Galaxy (e.g., Cayrel et al. 2004; Arnone et
al. 2005; Aoki et al. 2005, 2013; Cohen et al. 2004,2007,2008; Bonifacio
et al. 2009; Lai et al. 2009; Roederer 2009, Roederer et al. 2014).
Until recently, the numbers of stars with confirmed metallicities of
[Fe/H] $< -3.0$ was relatively small, making it difficult to infer the
global characteristics of EMP stars. Since 2005, more than 200 stars
with [Fe/H] $< -3.0$ have been confirmed, based on high-resolution
spectroscopic analyses (e.g., Barklem et al. 2005; Caffau et
al. 2013a,b; Yong et al. 2013), including more than 50 stars with [Fe/H] $<
-3.5$, and over 20 with [Fe/H] $< -4.0$ (Barklem et al. 2005; Frebel et
al. 2005, 2015; Cohen et al. 2008; Caffau et al. 2011a,b, 2012;
Bonifacio et al. 2012; Aoki et al. 2013; Caffau et al. 2013b; Spite et
al. 2013; Yong et al. 2013; Keller et al. 2014; Allende Prieto et al.
2015; Li et al. 2015; Placco et al 2015; Meléndez et al. 2016).

The SDSS database comprises over 900,000 stellar spectra, and is
now the dominant source of confirmed EMP stars. However, the relatively low resolution
($R \sim2000$) and limited signal-to-noise ratios (S/N $\sim30-40$)
of the SDSS spectra themselves make it difficult to derive elemental
abundances at very low metallicities. 

This paper is the third in a series devoted to analyses of Milky Way
halo stars based on the low-resolution SDSS spectroscopic data. We
present 107 stars with metallicities Fe/H] $ <-3.0$ for which the
stellar atmospheric parameters and chemical abundances of [Fe/H],
[Mg/H], and [Ca/H] have been estimated. We also report on a method to estimate
upper limits for these abundances from the SDSS spectra. This technique
enables a determination of the minimal S/N required to estimate the
abundance for a given element as a function of effective temperature
($T_{\rm eff}$) and surface gravity ($\log g$). 

This paper is outlined as follows. In Sect.~\ref{observations} we
briefly discuss the data used in the analysis, which is described in
Sect.~\ref{analysis}. In this section we also introduce our method to
estimate upper limits. Sect.~\ref{verification} reports on a
comparison with estimates from other analyses, including high-resolution
spectroscopy. The results are summarized in Sect.~\ref{results}.
Finally, we present our conclusions and a brief discussion in
Sect.~\ref{conclusions}.

\section{Observations}
\label{observations}

Our stellar spectra come from the SDSS. This project, started in 2000,
is now on its third extension, SDSS-IV (Alam et al. 2015), and comprises
a set of surveys devoted to a variety of areas, from cosmology to the
evolution of galaxies and the Milky Way to the search for extrasolar
planets. Its first extension, SDSS-II, included a specific stellar
project, the Sloan Extension for Galactic Understanding and Exploration
(SEGUE; Yanny et al. 2009), directed at investigation of the
structure, formation, and chemical evolution of the Galaxy. SEGUE-2, a
sub-survey of SDSS-III (Eisenstein et al. 2011), increased the number of stellar
spectra, and focused on observing distant halo stars. 

SEGUE, SEGUE-2, and other SDSS programs (including the main SDSS galaxy
redshift survey and BOSS, the Baryon Oscillations Spectroscopic Survey,
see Dawson et al. 2013) obtain spectra for colour-selected samples of
F-type main-sequence turn-off stars observed for calibration purposes.
Calibration stars taken during the main survey and BOSS have the
advantage of being located over the entire SDSS footprint at high
Galactic latitudes (rather than the limited number of directions probed
by SEGUE and SEGUE-2), and include halo stars at distances of up to
$\sim$100 kpc.

The spectra were obtained with a pair of spectrographs on the SDSS 2.5-m
telescope (Gunn et al. 2006; Smee et al. 2013) at Apache Point
Observatory, with a wavelength-dependent resolving power of $1300 < R <
3000$ over the spectral range $\sim 3800 < \lambda < 9000$\,{\AA}. The
spectrographs were updated before the beginning of BOSS observations, in
order to increase their efficiency and spectral range (to $3600 <
\lambda < 10000$\,{\AA}). Additional details on these spectra can be
found in the technical papers of the surveys (York et al. 2000; Yanny et
al 2009; Dawson et al. 2013; Alam et al. 2015), as well as in previous
papers in this series (Allende Prieto et al. 2014 and Fern\'andez-Alvar
et al. 2015; hereafter Paper~I and Paper~II).

\section{Analysis}
\label{analysis}

\subsection{Measurement of stellar parameters and chemical abundances}
\label{abundances}

\begin{figure*}
\centering
\includegraphics[scale=0.5]{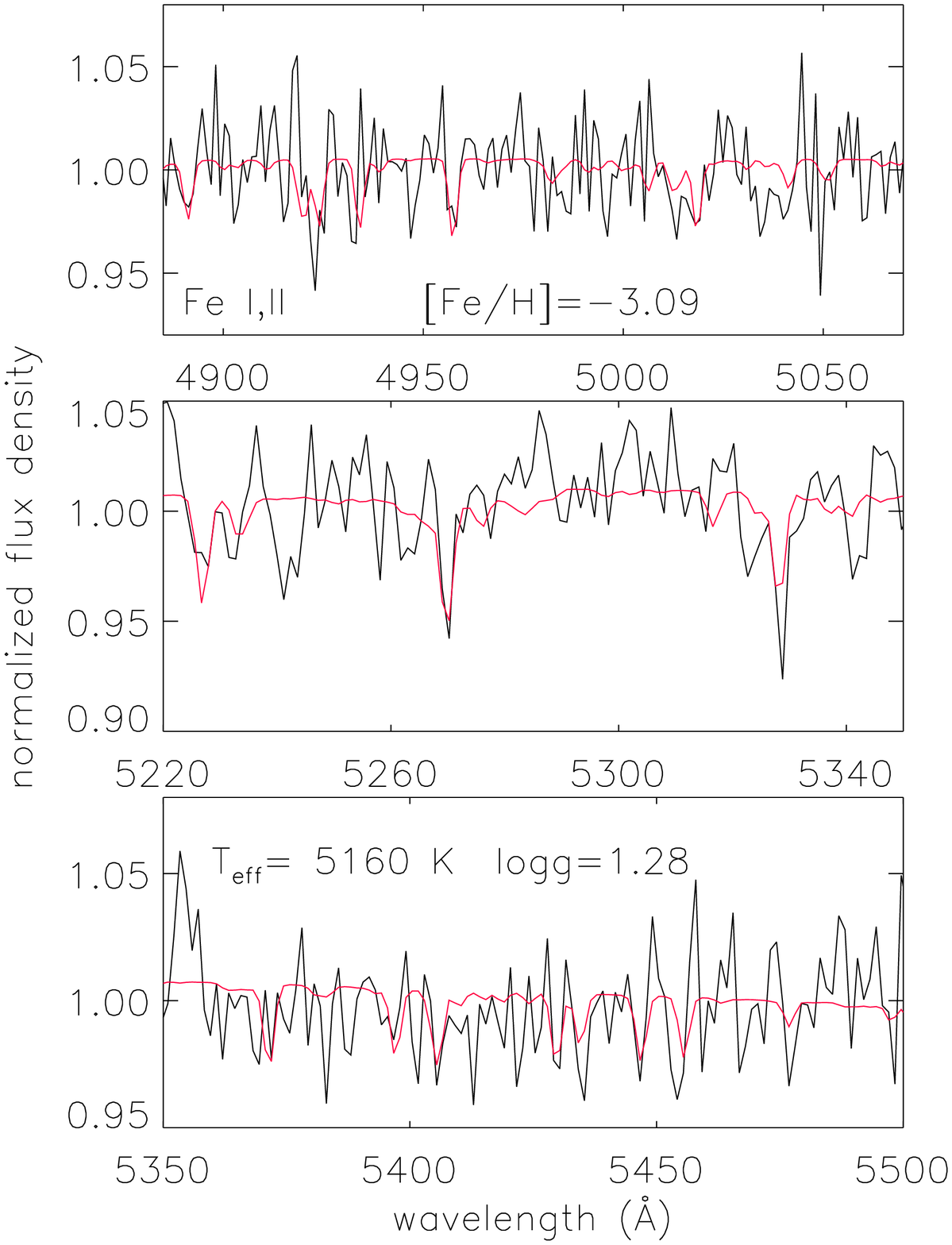}
\includegraphics[scale=0.5]{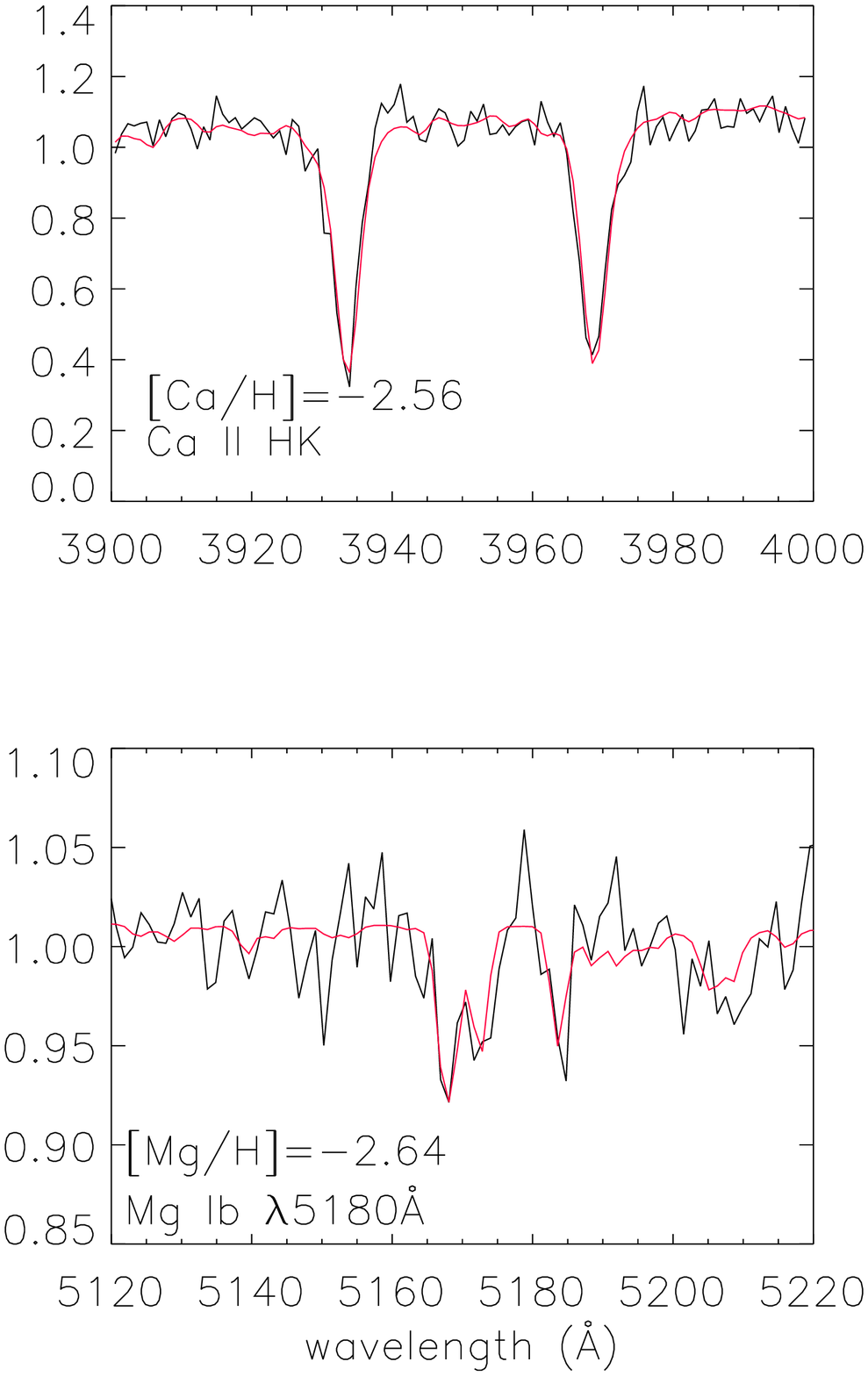}
\caption{Model fits for a SDSS/SEGUE star, SDSS J093339.24+310245.4, with the mean 
S/N of the sample, $\sim$40. The three panels in the left column
correspond to the fit regions from which the [Fe/H] abundance was
determined. The derived iron abundance is shown in the legend of
the upper panel, while the temperature and surface gravity estimates
are shown in the legend of the lower panel. The panels in the right
column show the CaII HK doublet and the
MgIb triplet that were fit to determine Ca and Mg abundances, with the
derived estimates shown in the legends.}
\label{cahmgh}
\end{figure*}

We wish to estimate stellar atmospheric parameters and individual element abundances
for a sample of extremely metal-poor stars belonging to the halo system.
Following the same strategy as described in Papers I and II, we made use
of the FERRE code (Allende Prieto et al. 2006) to constrain the stellar
atmospheric parameters -- effective temperature, $T_{\rm eff}$, surface
gravity, $\log g$, and the global metallicity, [M/H]\footnote{We adopt the
notation [X/H]$=\log_{10}\left(\frac{\rm N(X)}{\rm N(H)} \right)-\log_{10}\left(\frac{\rm N(X)}{\rm N(H)
} \right)_{\odot}$, where X is any chemical element, N(X) is the number
density of nuclei of this element, and N(H) is the number density of
hydrogen nuclei. [M/H] is the iron abundance determined from fitting the
available spectral range, which includes spectral features from other
metals.}. The search is performed by comparison with a grid of synthetic
spectra covering a wide range of parameter space, seeking the minimum
$\chi^{2}$ using quadratic Bezier interpolation between the model
spectra (Auer 2003). We employ the same 3-D spectral library ($T_{\rm eff}$,
$\log g$, and [M/H]) as in Paper~I, calculated from one-dimensional plane-parallel Kurucz model atmospheres (Castelli \& Kurucz 2003), which consider local thermodynamical equilibrium (LTE). The grid covers the ranges $4750< T_{\rm
eff} < 6500$~K, $0.5 < \log g < 4.5$, and $-5.0 <$ [M/H] $< +0.5$, in
steps of 250~K, 0.5 dex, and 0.5 dex, respectively. More details of the model
atmospheres, line data, and the opacities used in the generation of the
spectral library can be found in Paper I.  

The FERRE routine enables searches for one, several, or all of the
atmospheric parameters in a library of synthetic spectra. Once the
atmospheric parameters are determined, we search for a limited set of
elemental abundances, holding the atmospheric parameters fixed from the
analysis of the full spectrum. In our model grids we vary the abundances
of all metals relative to hydrogen in solar proportions (with the
exception of the $\alpha$-elements, which are enhanced by +0.4 dex for
metal-poor stars). Searching for [M/H], but restricting the fit to
regions dominated by individual lines of other elements, enables
estimation of the abundance of each of those elements, corresponding to
the value of [M/H] that best reproduces the shapes of their associated
lines. A more detailed explanation can be found in Paper~II. 

We first estimate the stellar parameters using the full spectral range
provided by the observations. The spectral range was limited to
$3850<\lambda<9190$\,{\AA} to ensure consistency with our previous
analysis in Papers~I and II, which examined both the BOSS spectra and
those from earlier SDSS/SEGUE observations. From this analysis we select
our primary targets to have metallicities in the range $-4.0 <$ [M/H] $<
-3.0$. We reject spectra that appear to be double-lined binaries or
white dwarfs. The S/N ratio (calculated as the median value per pixel in
the range $4885 < \lambda < 5500$\,{\AA} for each spectrum) varies
between 20 $<$ S/N $<$ 90; the derived values of $T_{\rm eff}$ and $\log
g$ varied between $5170 < T_{\rm eff} < 6500$~K and $0.5 < \log g <
4.5$, respectively. Holding $T_{\rm eff}$ and $\log g$ fixed, we repeat
the search in the [M/H] dimension of the grid by fitting selected
regions in the spectra that contain atomic lines of Fe, Ca, or Mg.

In this analysis we fit the following spectral
ranges (shown in Figure \ref{cahmgh}): $4885 < \lambda < 5070$\,{\AA},
$5220 < \lambda < 5280$\,{\AA}, $5295 < \lambda < 5500$\,{\AA} to
determine Fe abundances; $5160 < \lambda < 5190$\,{\AA} (the Mg Ib
triplet) for Mg abundances; and $3910 < \lambda < 3990$\, {\AA} (the
Ca~II H and K resonance doublet) for Ca abundances. These are the
regions with the highest sensitivity in the optical spectral range for
each element. The spectra were normalized by splitting them in 100
\AA\ pieces (200 \AA\ for the Fe I window between 4875 and 5510 \AA) and
dividing each piece by its mean flux value. Our synthetic spectra were treated in
the same manner. From the new metallicity estimates, we obtained the
abundance values of [Mg/H] and [Ca/H] ([Fe/H] is straightforward),
considering the relation with [Fe/H] adopted in the construction of the
spectral grid. 

\subsection{A method to estimate upper limits}
\label{upper}

\begin{figure}[!!h]
\centering
\includegraphics[scale=0.36, angle=90]{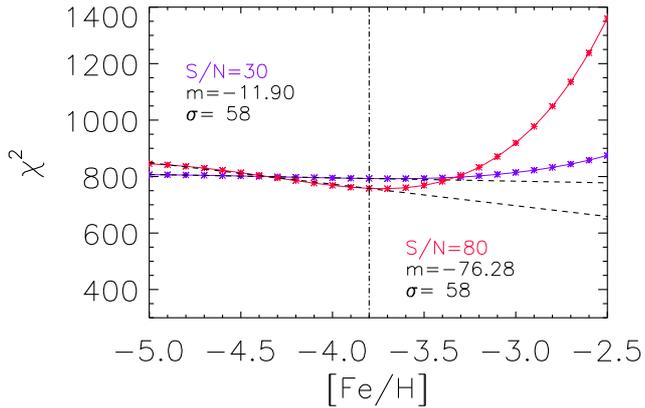}
\caption{The $\chi^{2}$ obtained from the evaluation of a simulated spectrum 
at [M/H]$_{0} = $[Fe/H]$_{0} = -3.8$ (marked with a vertical dotted line) compared with synthetic spectra over a range of
metallicities [M/H] (from $-5.0$ to $-2.5$ with a step of 0.1
dex). The blue line shows the resulting $\chi^{2}$ curve for a simulation with
S/N = 30, and the red line applies to S/N = 80. The two black dotted lines indicate the
linear fits up to the [M/H]$_{\rm 0}$ corresponding to the minimum
$\chi^{2}$, from which the slope and its uncertainty are derived.}
\label{method}
\end{figure}

\begin{figure*}[!!h]
\centering
\includegraphics[width=6cm]{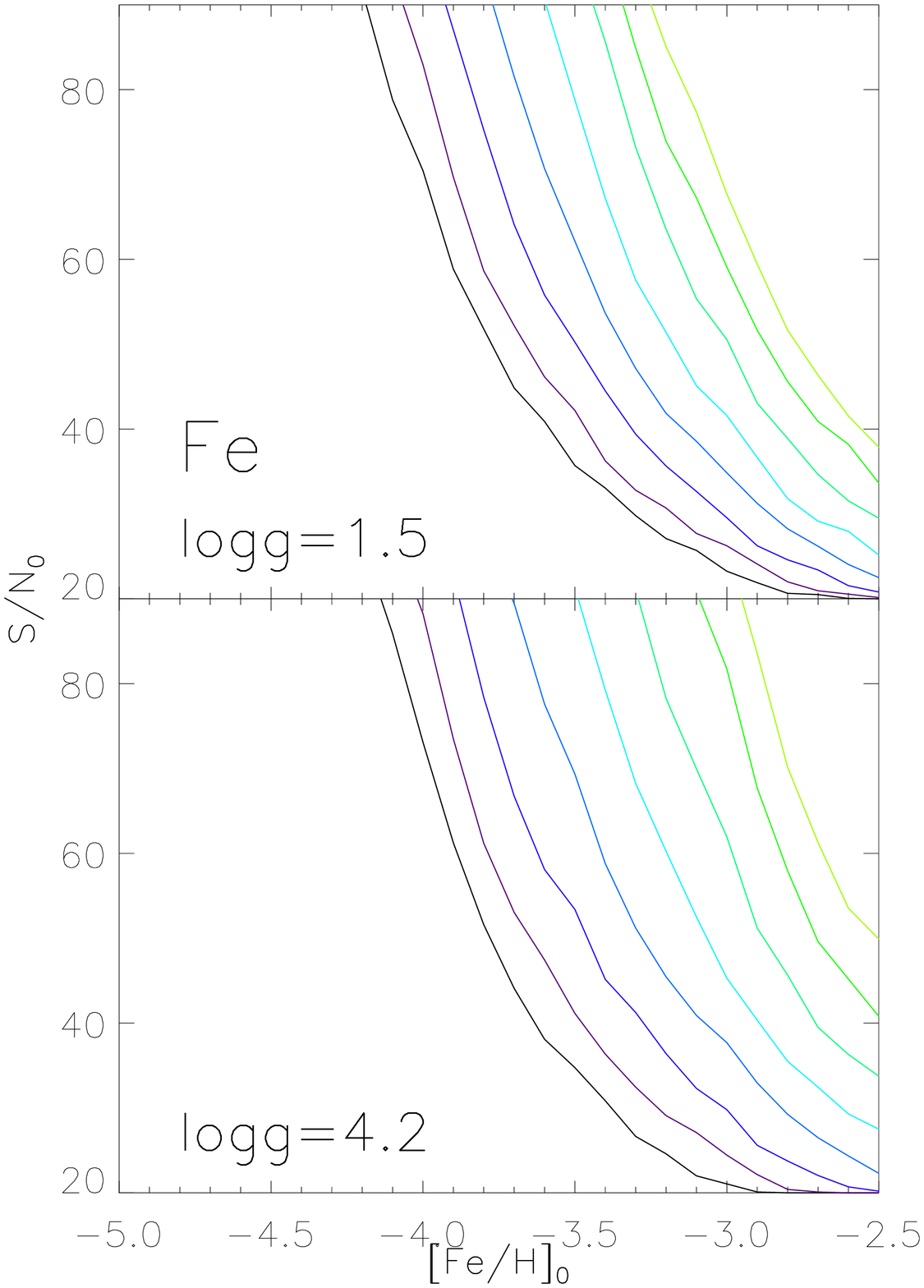}
\includegraphics[width=6cm]{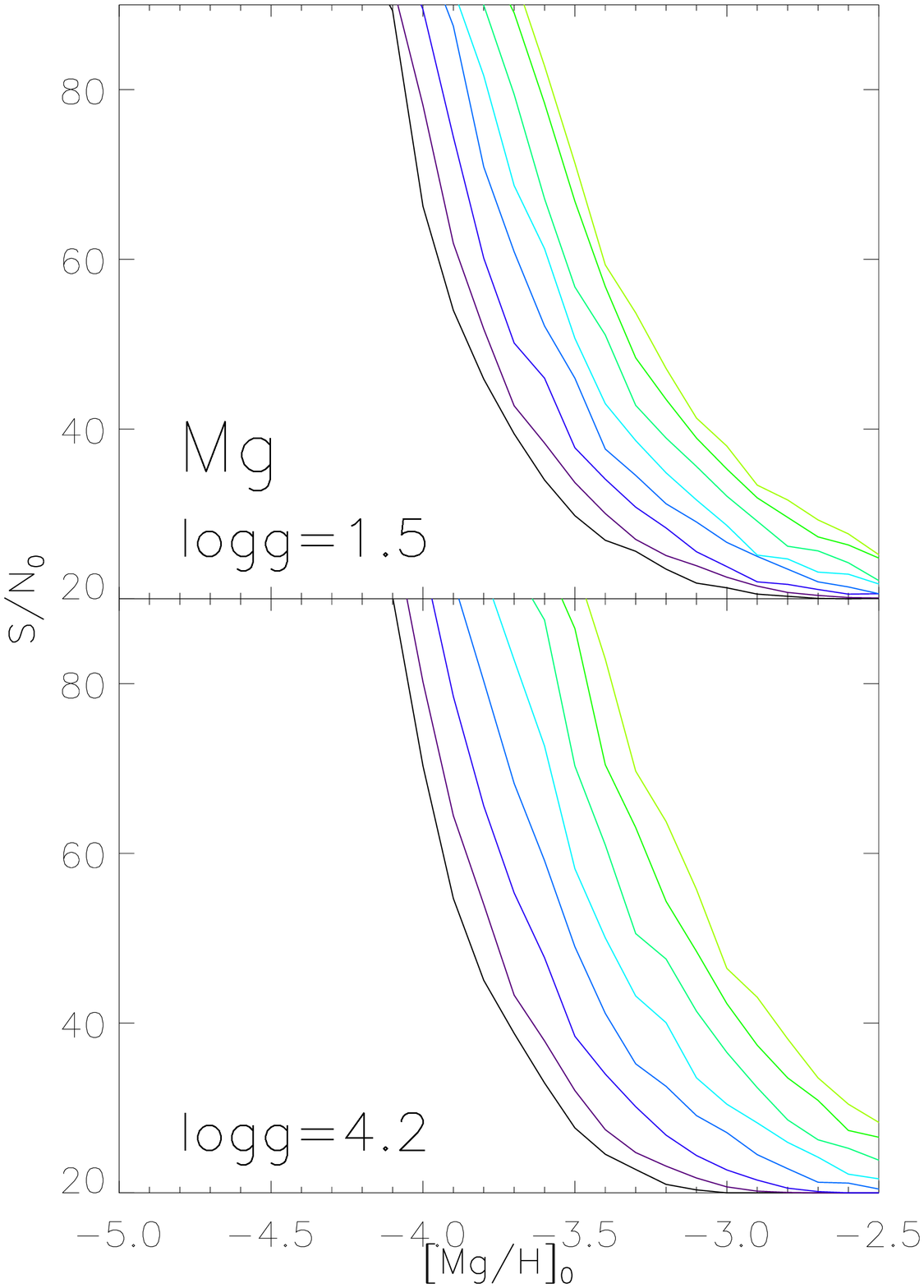}
\includegraphics[width=6cm]{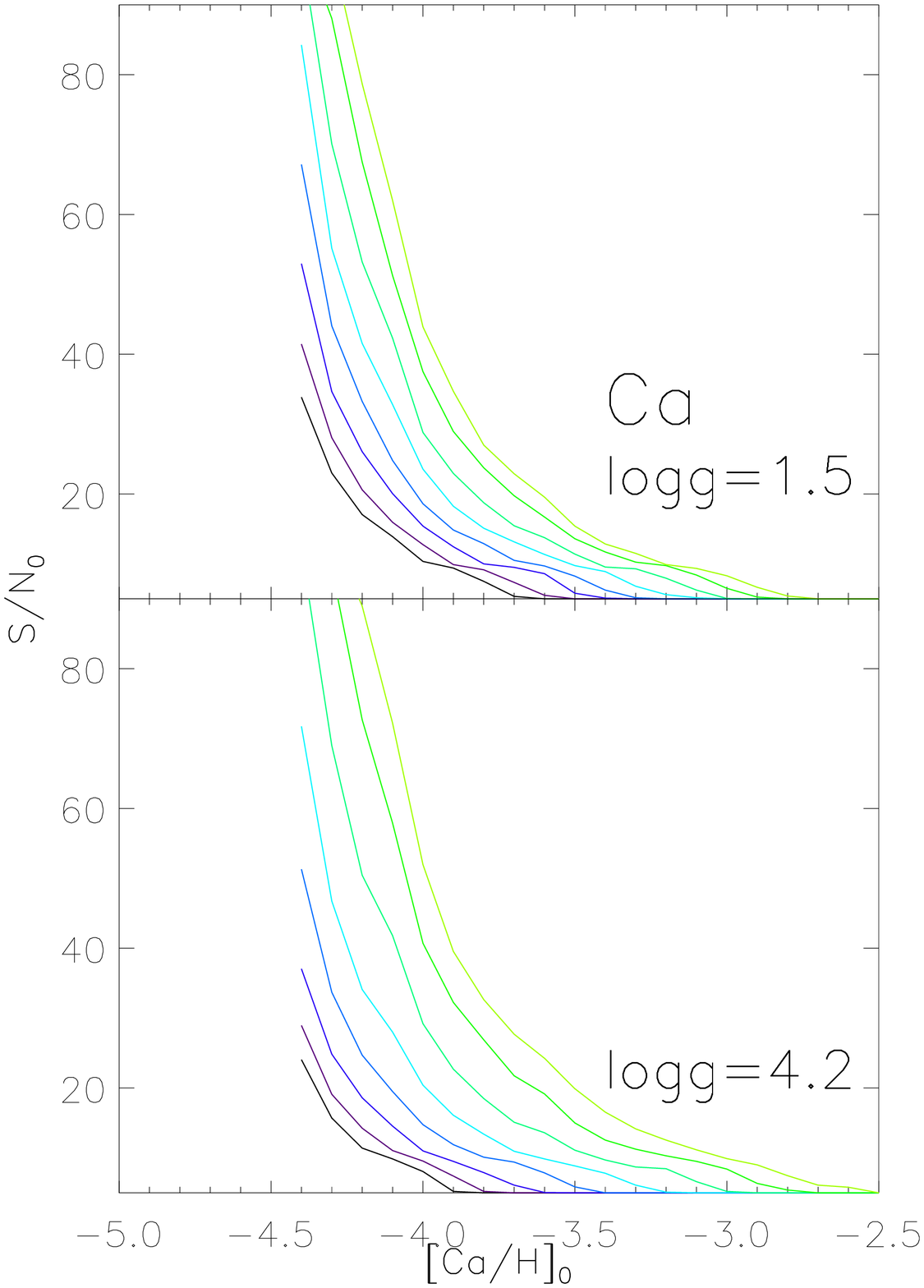}
\caption{Minimum signal-to-noise values, S/N$_{\rm 0}$, required to
reliably determine [Fe/H]$_{\rm 0}$, [Mg/H]$_{\rm 0}$ and [Ca/H]$_{\rm
0}$ minimum abundances, evaluated at $\log g = 1.5$ and $\log g=4.5$ and
4750$ < T_{\rm eff} <$ 6500~K (darker colors correspond to
lower temperatures). The increase of $T_{\rm eff}$ complicates the
[Fe/H] determination, mainly in the case of Fe and Mg, and a higher
S/N is necessary.  The sensitivity of the Ca II HK
resonance lines makes it possible to determine [Ca/H] at very low abundances.}
\label{hl_femg}
\end{figure*}

The relatively low S/N and modest resolution of our data complicate the
estimation of some elemental abundances, in particular for Fe and Mg,
whose line detections become marginal for low metallicities at
the S/N of our spectra. For this reason, we developed a methodology to
determine upper limits on the abundance of an element as a function of
S/N, $T_{\rm eff}$, $\log g$, and metallicity, [M/H]. 

We simulated the observed spectra by smoothing our spectral library to
$R = 2000$ and adding Gaussian noise. From a simulated spectrum with a
particular S/N, $T_{\rm eff}$, $\log g$, and [M/H] (hereafter
[M/H]$_{\rm 0}$), we evaluated the $\chi^{2}$ and its error ($\sigma =
\sqrt{4\chi^{2}}$), comparing with noise-free spectra over $-5.0 <
$[M/H]$ < -2.5$ for the same spectral windows used to determine
the abundances of Fe, Mg, and Ca. We calculated the slope of the
$\chi^{2}$ curve in the range $-5.0 <$ [M/H] $<$ [M/H]$_{\rm 0}$.
Figure~\ref{method} illustrates the methodology. Repeating this process
at different S/N ratios, from 20 to 90 (in steps of 5), we defined the
minimum S/N (hereafter S/N$_{\rm 0}$) at which the slope
becomes significant (i.e., its error is lower than its value) for each
[M/H]$_{\rm 0}$. This calculation was performed a hundred times to
statistically refine this minimum S/N$_{\rm 0}$ value. 

Evaluating simulations at $T_{\rm eff}$ values from 4750 to 6500~K in
steps of 250~K, two $\log g$ values, 1.5 and 4.2, and metallicities
from $-4.8$ to $-2.5$ in steps of 0.1 dex, we established the minimum
S/N$_{\rm 0}$ vs. [Fe/H]$_{\rm 0}$ limit curves for each combination of
stellar atmospheric parameters. The S/N range considered was 20-90 in
the cases of Fe and Mg. We extended the lower limit down to S/N = 5 in
the Ca evaluation, since the S/N over 3900$ < \lambda < 4000$\,{\AA} in the
analysed spectra is lower than 20 in some cases. These curves were used
to evaluate the reliability of our Fe, Mg, and Ca abundance
determinations. Figure \ref{hl_femg} shows the results for the case of
$\log g = 1.5$ and 4.2. We accepted a given abundance determination when
it was higher than the corresponding S/N at which the modulus of the
difference with the S/N of the observation is smallest. Otherwise, if
the value determined by FERRE is lower than this limit, we considered
the estimate as an upper limit for the abundance. 

This method can be used for the analysis of low-resolution, low-S/N
spectra. From Figure \ref{hl_femg} we can infer that $T_{\rm eff}$
impacts determination of the abundances of the three elements considered
more than $\log g$, as expected.

Iron is the element that is most difficult to measure. Although
the number of available lines is larger than for the other two elements,
their weakness in this metallicity regime makes them challenging to
measure in the presence of noise. Our calculations indicate that a S/N
$> 90$ is necessary to reliably determine [Fe/H] $\le -4.2$ over the
$T_{\rm eff}$ and $\log g$ ranges considered and our selected
spectral range. At S/N $ \sim$90 ,the upper limit changes by $\sim$1
dex, from $-4.0$ to $-3.0$, between 4750 to 6500 K at $\log g$ = 1.5
(becoming larger as $\log g$ increases). By contrast, the calcium
abundance can be measured down to [Ca/H] $\sim -4.2$ for spectra with a
median S/N $\sim$ 10 for stars with $T_{\rm eff}$ = 4750~K and $\log g =
4.2$.

In order to provide a convenient tool to estimate the minimum S/N
required to measure the abundances of these elements from the spectral windows considered in this work, we derived an
analytic function that fits the curves obtained from the simulations. We
found that these curves can be well-reproduced by a second-order
polynomial (a convex parabola) and a straight line through the vertex of
the parabola.

$$ {\rm S/N} = 
\begin{cases}
a(T_{\rm eff}) {\rm [X/H]}^{2} + b(T_{\rm eff}){\rm [X/H]} + c(T_{\rm eff}) & \text{if [X/H] $< \frac{-b}{2a}$}\\
d(T_{\rm eff}){\rm [X/H]} + e(T_{\rm eff}) & \text{if [X/H] $> \frac{-b}{2a}$}
\end{cases}
$$

We model the dependence of the polynomial coefficients with $T_{\rm
eff}$ as second-order polynomials. The domain of applicability for each
second-order polynomial is defined below and above certain $T_{\rm eff}$
values (5500~K at $\log g = 1.5$ and 6000~K at $\log g = 4.2$). 

   \begin{equation}
      a(T_{\rm eff}) = p_{11}T_{\rm eff}^{2} + p_{12}T_{\rm eff} + p_{13}
   \end{equation}
   \begin{equation}
      b(T_{\rm eff}) = p_{21}T_{\rm eff}^{2} + p_{22}T_{\rm eff} + p_{23}
   \end{equation}
   \begin{equation}
      c(T_{\rm eff}) = p_{31}T_{\rm eff}^{2} + p_{32}T_{\rm eff} + p_{33}
   \end{equation}

A single first-order polynomial well-fits the dependence of the other
two coefficients on $T_{\rm eff}$, 

   \begin{equation}
      d(T_{\rm eff}) = l_{11}T_{\rm eff} + l_{12}
   \end{equation}
   
   \begin{equation}
      e(T_{\rm eff}) = l_{21}T_{\rm eff} + l_{22}
   \end{equation}

Tables \ref{tb-coef1}, \ref{tb-coef2}, and \ref{tb-coef3} show the
coefficients of these polynomials for each element for the two $\log g$
values evaluated, 1.5 and 4.2.

\begin{table}
\caption{\label{tb-coef1} Second-order polynomial coefficients that
reproduce the parabolic and linear polynomial parameters,
as a function of $T_{\rm eff}$, for each of the $\log g$ values
considered, 1.5 and 4.2, for estimation of Fe.}
\center
\begin{tabular}{crrr}
\hline\hline

 Fe & $\log g = 1.5$ & & \\
 \hline\hline
 & $T_{\rm eff} < 5500$ K & & \\
 \hline
 $p_{1}$ & $-$1.122267e$-$04             &               1.116517e+00            &               $-$2.725528e+03 \\
 $p_{2}$ & $-$7.752576e$-$04             &               7.661329e+00            &               $-$1.862050e+04 \\
 $p_{3}$ & $-$1.307124e$-$03             &               1.285527e+01            &               $-$3.112059e+04 \\
 \hline
 & $T_{\rm eff} > 5500$ K & & \\
 \hline
 $p_{1}$ & $-$4.702741e$-$05             &               5.608045e$-$01          &               $-$1.619006e+03 \\
 $p_{2}$ & $-$2.673086e$-$04             &               3.143082e+00            &               $-$8.982544e+03 \\
 $p_{3}$ & $-$3.688757e$-$04             &               4.301038e+00            &               $-$1.219482e+04 \\
\hline\hline
 $l_{1}$ & $-$1.493794e$-$02             &               6.539818e+01 & \\
 $l_{2}$ & $-$3.479122e$-$02             &               1.706138e+02 & \\
 \hline\hline
 Fe & $\log g = 4.2$ & & \\
\hline\hline
 & $T_{\rm eff} < 6000$ K & & \\
\hline
 $p_{1}$ &  6.350261e$-$05               &                $-$6.807567e$-$01       &               1.868088e+03  \\
 $p_{2}$ &  4.025916e$-$04               &                $-$4.391630e+00         &               1.219878e+04  \\
 $p_{3}$ &  6.643223e$-$04               &                $-$7.321549e+00         &               2.047375e+04  \\
 \hline
 & $T_{\rm eff} > 6000$ K & & \\
 \hline
 $p_{1}$ & 4.131073e$-$04                &               $-$4.953010e+00          &               1.490776e+04  \\
 $p_{2}$ & 2.245832e$-$03                &               $-$2.701466e+01          &               8.152300e+04  \\
 $p_{3}$ & 3.071746e$-$03                &               $-$3.703024e+01          &               1.119596e+05  \\
\hline\hline
 $l_{1}$ & $-$5.311476e$-$03             &               1.899590e+01  & \\
 $l_{2}$ & $-$7.989891e$-$03             &               3.985540e+01  & \\
 \hline\hline
\end{tabular}
\end{table} 
 
\begin{table}
\caption{\label{tb-coef2} The same as in Table \ref{tb-coef1}, but for
estimation of Mg.}
\center
\begin{tabular}{crrr}
\hline\hline 
  Mg & $\log g = 1.5$ & & \\
\hline\hline
 & $T_{\rm eff} < 5500$ K & & \\
\hline
 $p_{1}$ & 1.205388e$-$04             &              $-$1.318294e+00             &               3.680773e+03 \\
 $p_{2}$ & 9.044162e$-$04             &              $-$9.930854e+00             &               2.772508e+04 \\
 $p_{3}$ & 1.706439e$-$03             &              $-$1.876632e+01             &               5.232795e+04 \\
 \hline
 & $T_{\rm eff} > 5500$ K & & \\
 \hline
 $p_{1}$ & 4.551823e$-$05             &              $-$5.582935e$-$01           &               1.766349e+03 \\
 $p_{2}$ & 3.205082e$-$04             &              $-$3.960549e+00             &               1.252583e+04 \\
 $p_{3}$ & 5.621326e$-$04             &              $-$6.969943e+00             &               2.201842e+04 \\
\hline\hline
 $l_{1}$ & $-$2.316824e$-$03          &                4.812897e+00 & \\
 $l_{2}$ &   $-$1.307164e$-$03        &                8.809123e+00 & \\
 \hline\hline
  Mg & $\log g = 4.2$ & & \\
\hline\hline
 & $T_{\rm eff} < 6000$ K & & \\
\hline
 $p_{1}$ & 8.279952e$-$05             &              $-$9.399753e$-$01           &                2.729978e+03 \\
 $p_{2}$ &  6.308861e$-$04            &               $-$7.190066e+00            &                2.083195e+04 \\
 $p_{3}$ &  1.210125e$-$03            &               $-$1.378693e+01            &                3.977522e+04 \\
 \hline
 & $T_{\rm eff} > 6000$ K & & \\
 \hline
 $p_{1}$ & 6.482425e$-$06             &              $-$1.038303e$-$01           &               4.460068e+02 \\
 $p_{2}$ & 5.010227e$-$05             &              $-$8.068283e$-$01           &               3.330069e+03 \\
 $p_{3}$ & 1.011057e$-$04             &              $-$1.570228e+00             &               6.187633e+03 \\
\hline\hline
 $l_{1}$ &  $-$9.710505e$-$04         &                $-$3.277127e+00 \\
 $l_{2}$ &   4.284561e$-$03           &                $-$2.491109e+01 \\
 \hline\hline
\end{tabular}
\end{table}

\begin{table}
\caption{\label{tb-coef3} The same as in table \ref{tb-coef1}, but for
estimation of Ca.}
\center
\begin{tabular}{crrr}
\hline\hline 
  Ca & $\log g = 1.5$ & & \\
\hline\hline
 & $T_{\rm eff} < 5500$ K & & \\
\hline
 $p_{1}$ & $-$3.711640e$-$04           &               3.904669e+00             &              $-$1.013811e+04 \\
 $p_{2}$ &  $-$3.092381e$-$03          &               3.245765e+01             &              $-$8.415880e+04 \\
 $p_{3}$ & $-$6.420058e$-$03           &               6.725487e+01             &              $-$1.741799e+05 \\
 \hline
 & $T_{\rm eff} > 5500$ K & & \\
 \hline
 $p_{1}$ & 3.881031e$-$06              &             $-$9.518567e$-$02          &              5.899625e+02 \\
 $p_{2}$ & 2.968447e$-$05              &             $-$8.304656e$-$01          &              5.092121e+03 \\
 $p_{3}$ & 6.847159e$-$05              &             $-$1.908617e+00            &              1.119159e+04 \\
\hline\hline
 $l_{1}$ & 2.485399e$-$04              &             $-$1.451917e+01  \\
 $l_{2}$ &  7.746718e$-$03             &              $-$8.033748e+01 \\
 \hline\hline
  Ca & $\log g = 4.2$ & & \\
\hline\hline
 & $T_{\rm eff} < 6000$ K & & \\
\hline
 $p_{1}$ &   5.989651e$-$05            &               $-$4.844238e$-$01        &                9.931608e+02 \\
 $p_{2}$ &  4.289678e$-$04             &              $-$3.375793e+00           &                6.682556e+03 \\
 $p_{3}$ &   7.663057e$-$04            &               $-$5.832676e+00          &                1.102958e+04 \\
 \hline
 & $T_{\rm eff} > 6000$ K & & \\
 \hline
 $p_{1}$ & 2.605201e$-$05              &             $-$2.514639e$-$01          &               6.919529e+02 \\
 $p_{2}$ & 2.809412e$-$04              &             $-$3.033114e+00            &               8.964729e+03 \\
 $p_{3}$ & 7.024900e$-$04              &             $-$8.000266e+00            &               2.432530e+04 \\
\hline\hline
 $l_{1}$ &   $-$1.835714e$-$03         &              $-$1.504880e+00 \\ 
 $l_{2}$ &  4.354132e$-$03             &              $-$5.797910e+01 \\
 \hline\hline
\end{tabular}
\end{table}

\section{Verification with other analyses}
\label{verification}
\subsection{Comparison with the SSPP analysis}
\label{sspp}

The SEGUE Stellar Parameter Pipeline (SSPP) was developed to analyse the
stellar spectra gathered in the SDSS/SEGUE surveys (see Lee et al.
2008a,b; Allende Prieto et al. 2008; Smolinksi et al. 2011; Lee et al.
2011 for details), and can be used to estimate $T_{\rm eff}$, $\log g$,
[Fe/H], and [$\alpha$/Fe]\footnote{The [$\alpha$/Fe] corresponds to
the [Mg/Fe], [Si/Fe], [Ca/Fe], and [Ti/Fe] global measurement, from the
spectral range 4500 -- 5500\,{\AA} fit to the observational data with synthetic
spectra.}. Recently, the SSPP has been extended to be capable of
estimating [C/Fe] as well (Lee et al. 2013). For the sake of comparison,
we have also evaluated our sample of moderate-resolution SDSS spectra
with this tool.

We compare the stellar parameters and abundance measurements obtained
with the SSPP compared with our present analysis results. Our analysis
returns a lower estimated $T_{\rm eff}$ with respect to the SSPP
($\delta = -353$~K), and a modest dispersion ($\sigma =$ 277~K). This is
not unexpected, since our method is purely spectroscopic, and the SSPP
uses a combination of photometric and spectroscopic techniques. It has
long been recognized that spectroscopically-determined temperature
estimates can be up to several hundred Kelvin cooler than photometric
estimates. Regarding the $\log g$ estimate, a large systematic deviation
exists for stars for which we obtain surface gravity estimates of $\log
g <$ 3 ($\delta = -1.26$~dex, $\sigma =$ 1.56~dex). This is again not
surprising, since estimates of surface gravity are particularly
challenging at low metallicity from low-S/N spectra. The [Fe/H]
comparison exhibits a relatively small negative offset $\delta =
-0.16$~dex, with a dispersion $\sigma =$ 0.33~dex. Essentially all of this offset can be accounted for by the differences in the
temperature estimates. 

Finally, we compare the SSPP estimate of [$\alpha$/Fe] with our [Ca/Fe]
and [Mg/Fe] measurements. Both show an offset $\delta\sim -0.15$~dex and
dispersion $\sigma\sim$ 0.26~dex. Stars with the highest and lowest
[Ca/Fe] and [Mg/Fe] estimates from our own estimates exhibit larger
differences with respect to the SSPP results.

In order to clarify whether the differences in the estimated stellar
parameters $T_{\rm eff}$ and $\log g$ severely impact the resulting
abundance measurements, we repeat the [Fe/H], [Ca/Fe], and [Mg/Fe]
determinations with FERRE, but after replacing the stellar parameters
with those from SSPP. The resulting offset in the [Fe/H] determination
compares better with the SSPP [Fe/H], although the dispersion increases
slightly ($\delta = -0.09$~dex, $\sigma =$ 0.38~dex). Conversely, the
dispersion in the comparison of the SSPP [$\alpha$/Fe] with the new
[Ca/Fe] and [Mg/Fe] estimates increases by more than 0.2~dex, although
the offset in the case of [Mg/Fe] is reduced from $-0.14$~dex to
$-0.01$~dex. Thus, differences in Teff and log g are not solely
responsible for the derived abundance contrast with the SSPP.

\subsection{Comparison with high-resolution analyses}
\label{highres}

There are several studies of EMP candidates from SDSS/SEGUE that have
been followed-up and analysed with high-resolution spectra reported in
the literature. Here we consider a comparison of our present results 
with these measurements.

SDSS J031745.82+002304.1 was analysed by Bonifacio et al. (2012) as one
of their 16 EMP candidates found in the SDSS/SEGUE database. Our
measurements for this star are in very good agreement with their
estimates from high-resolution spectra. We obtain $T_{\rm eff} =$
5780~K, $\log g$ = 3.72, [Fe/H] $> -3.40$, [Ca/Fe] $\geq +0.62$ and
[Mg/Fe] $\geq +0.21$, while they obtained $T_{\rm eff} =$ 5786~K, $\log
g$ = 4.02, [Fe/H] $= -3.46$, [Ca/Fe] $= +0.75$ ($+0.60$ from Ca~I lines)
and [Mg/Fe] $= +0.38$. 

Aoki et al. (2013) determined $T_{\rm eff}$, $\log g$, [Fe/H], [Ca/Fe],
and [Mg/Fe] for 70 VMP and EMP stars selected from SDSS/SEQUE. Nine of these objects
are in common with our sample: [Fe/H] measurements from both analyses
are available for 7 stars, [Ca/Fe] for 5 stars, and [Mg/Fe] for 6 stars.
Our $T_{\rm eff}$ estimates are offset by $-241$~K from the Aoki et al.
results, with a dispersion $\sigma =$ 169~K. The $\log g$ comparison
exhibits a larger offset and dispersion, $\delta = -1.03$~dex and
$\sigma = 1.21$ dex, similar to the comparison obtained with respect to
the SSPP results. The [Fe/H] estimates compare reasonably well, with an
offset of $\delta = -0.12$~dex and $\sigma = 0.21$~dex. The [Mg/Fe]
results are only in fair agreement, $\delta = -0.08$~dex and $\sigma =
0.30$~dex. In contrast, our [Ca/Fe] estimates are 0.43~dex higher, with
a dispersion of $\sigma = 0.39$~dex. However, the Aoki et al.
determination of [Ca/H] in some of their stars came from the Ca~I 4226
\AA\ equivalent width, which is more sensitive to non-LTE effects than
the Ca~II HK doublet we have considered.

We identify SDSS J132250.59+012342.9 as an EMP star, as previously
reported by Placco et al. (2015). For this star we obtained the
following parameters: $T_{\rm eff} =$ 5234~K, $\log g =$ 0.84, [Fe/H] $
= -3.32$, [Ca/Fe] $= +0.23$ and [Mg/Fe] $= +0.30$. Placco et al.
obtained a lower $T_{\rm eff}$ = 5008 $\pm$ 100~K, a higher $\log g =
1.95 \pm 0.20$, and a lower [Fe/H] $= -3.64 \pm 0.05$, which are in
reasonable agreement with our estimates. They also obtained [Ca/Fe]$ =
+0.23 \pm 0.08$ and [Mg/Fe] = $+0.25 \pm 0.05$, which are also in
good agreement with our results.

Recently, Susmitha Rani et al. (2016) reported [Fe/H] $= -3.42 \pm 0.19$ for the
star SDSS J134338.67+484426.6. This star is also included in our sample;
we obtained estimates of $T_{\rm eff} =$ 5307~K, $\log g =$ 0.51, [Fe/H]
$< -3.7$, [Ca/H] $= -3.40$, and [Mg/H] $= -3.52$. Their temperature
estimate is higher than ours, $T_{\rm eff} =$ 5620~K, and their $\log g
=$ 3.44 is considerably larger. We only determined an [Fe/H] upper
limit, which is lower than their [Fe/H] measurement. They also
determined [Ca/H] $= -3.23 \pm 0.16$, which is consistent with our
estimate, and [Mg/H] $= -3.27 \pm 0.16$, slightly lower than our
estimate. 

In conclusion, the several comparisons we have performed reveal that our
temperature estimates are generally lower than previously reported
results, by about 200-250~K, due to our use of spectroscopic estimates,
a significant under-estimate in our determination of $\log g$, a slight
under-estimate of [Fe/H] (understandable from the $T_{\rm eff}$ offset)
and a dispersion $\sim$0.3~dex in the derived [Ca/Fe] and [Mg/Fe]. 


\section{Results}
\label{results}

Our final sample of EMP stars is listed in Table \ref{SEGUE_coor},
along with their $ugriz$ magnitudes and heliocentric velocities,
calculated by the SSPP (the typical accuracy for these velocities is
on the order of 5 km/s). Tables \ref{SEGUE_sp} and \ref{SEGUE_ab} list our estimated stellar
atmospheric parameters, $T_{\rm eff}, \log g$, and [M/H], and the
derived abundances [Fe/H], [Ca/H], and [Mg/H], as well as their
uncertainties. The minimum $\chi^{2}$ searches with FERRE were repeated 10 times, with added random noise. The uncertainties are estimated from the standard deviation.  We only list stars for which we obtained $-4.0 <$ [Fe/H]
$< -3.0$. In cases for which our estimates were lower than the upper
limit associated with the stellar parameters and S/N of the
spectrum from which it was measured, a corresponding upper limit is
stated. We obtained [Fe/H] measurements for 44 SDSS/SEGUE spectra and
4 BOSS spectra, 48\% and 24\% of each sample, respectively. In the case
of [Mg/H], we determined reliable estimates for 86 (93\%) and 13 (76\%)
of the stellar spectra in the SDSS/SEGUE and BOSS samples, respectively.
The [Ca/H] estimates were all reliable measurements. As noted above, Fe
is the most difficult abundance to measure due to the weakness of its
lines.

\subsection{[Ca/Fe] and [Mg/Fe]}
\begin{figure}
\center
\includegraphics[scale=0.37, angle=90]{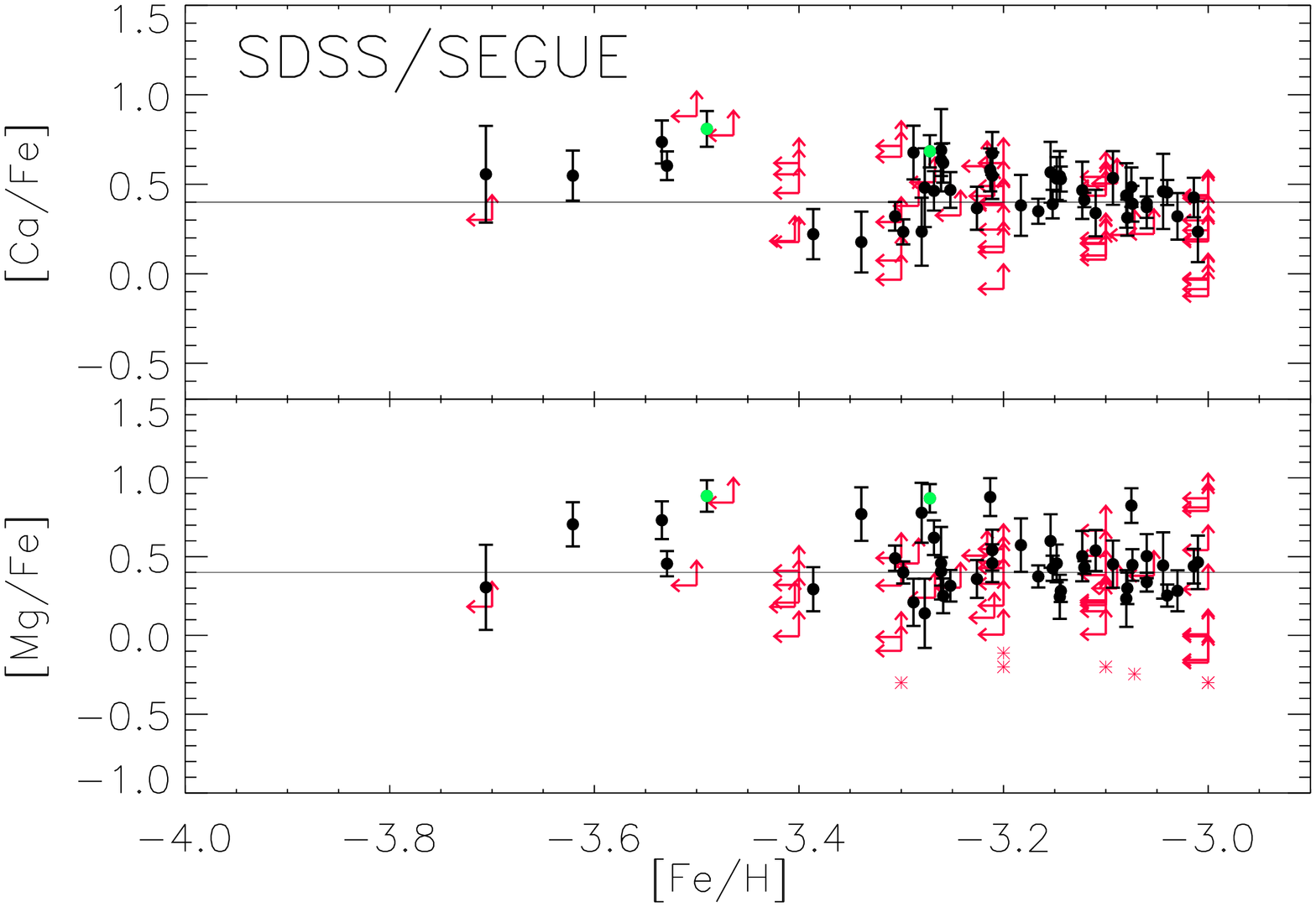}
\includegraphics[scale=0.37, angle=90]{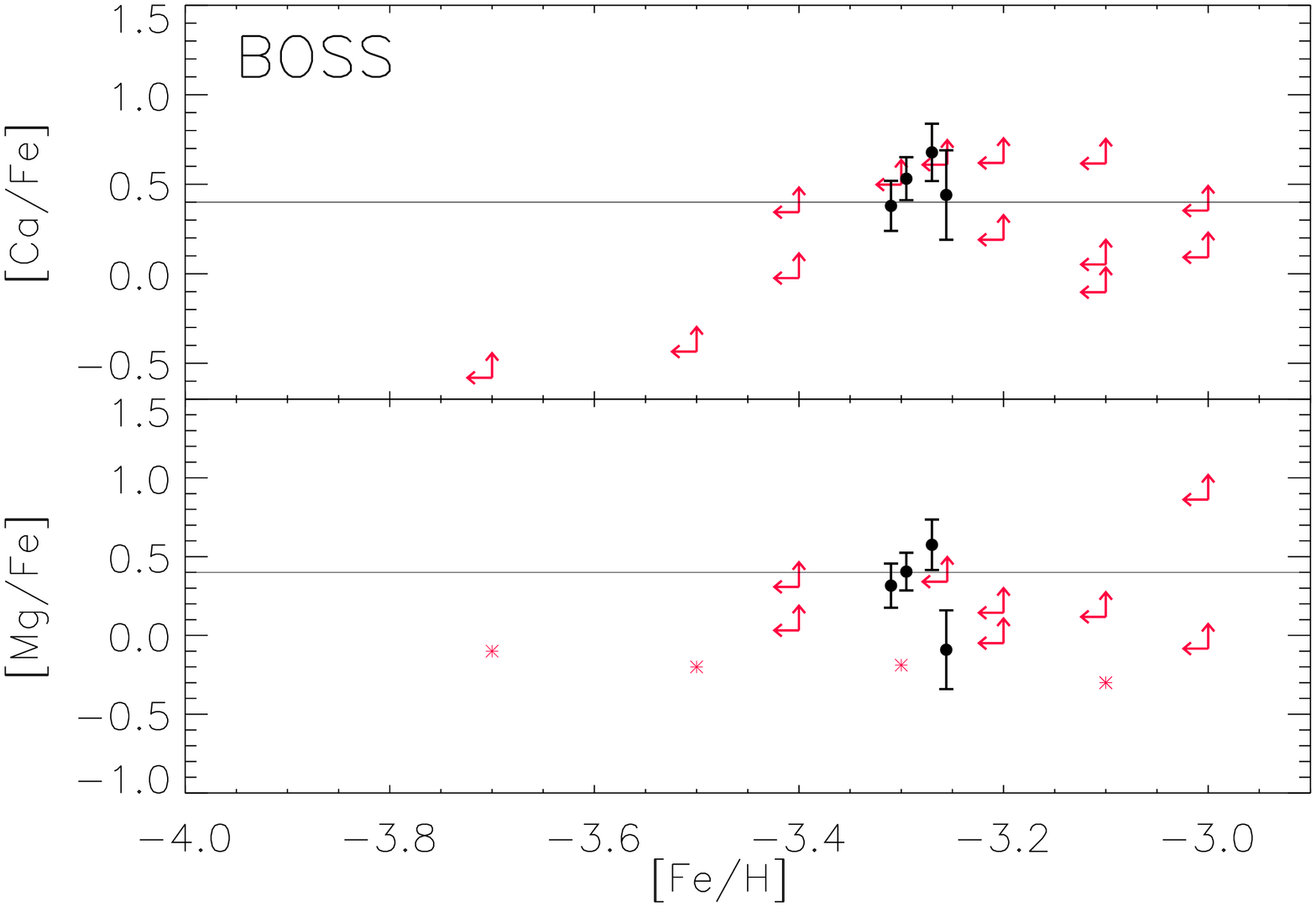} 
\caption{[Ca/Fe] and [Mg/Fe] ratios, as a function of [Fe/H], from the analysis
of our SDSS/SEGUE and BOSS samples. The reliable estimates after
applying the upper limit evaluation are indicated as black dots. Red
arrows show the cases where only an upper limit for the abundances could
be obtained, and the direction the ratio would be situated in this
diagram.  The two stars for which we obtain high [Ca/Fe] and [Mg/Fe] are
shown as green dots.}
\label{cafe}
\end{figure}

The $\alpha$-element enrichment is of particular interest in this
extremely low-metallicity regime, since it can provide valuable
information on the nucleosynthesis histories of the first generations of
stars. Figure~\ref{cafe} shows our derived [Ca/Fe] and [Mg/Fe] (split
into SDSS/SEGUE and BOSS stars in the top and bottom panels) as a
function of [Fe/H]. We use arrows to indicate upper limits on [Fe/H]
(hence lower limits on [Ca(Mg)/Fe]). In cases for which neither
Ca (or Mg) nor Fe abundances were determined, we plot their
ratio with an asterisk. Overall, it appears that most of the stars present
a value of [Ca(Mg)/Fe] consistent with the expected halo-star value of
$\sim$+0.4. From inspection of this figure, it is apparent that, at
the lowest metallicities, there are more stars above [$\alpha$/Fe]$
=+0.4$ than below this value. 

From the stars for which we obtained reliable determinations of the
abundances, we took into account results for which the deviation from
[Ca/Fe] = $+0.4$ was higher than three times the uncertainties of the
ratios over the metallicity, or the iron abundance, in each case. We
found two stars that exhibit ratios of [Ca/Fe] and [Mg/Fe] significantly
higher than $+0.4$ among the pre-BOSS stars: SDSS J035622.42+114705.4
([Ca/Fe] $= +0.68 \pm 0.09$; [Mg/Fe] $= +0.87 \pm 0.11$), and SDSS
J031259.10-061957.1 ([Ca/Fe] $= +0.89 \pm 0.10$; [Mg/Fe] $ = +0.89 \pm
0.14$).
These stars are shown in Figure~\ref{cafe} as green dots. However, note
that the SSPP estimates for these stars are [$\alpha$/Fe]$ = +0.14$ and
+0.41, respectively. Hence, more accurate chemical-abundance
estimates are required in order to confirm their status as
$\alpha$-enhanced stars.

\subsection{[C/Fe]}

The SSPP provides [C/Fe] measurements for 57 stars in our sample with
detected carbon. After examining the resulting values, we find 28
carbon-enhanced metal-poor (CEMP) stars ([C/Fe] $>$ +0.7), from which we obtain a cumulative frequency of CEMP stars below [Fe/H] = -3.0 of $\sim26\%$. We also verify that the frequency of CEMP stars increases as
the metallicity decreases. Dividing our sample at the median metallicity
for the stars with measured metallicity ([Fe/H] $= -3.2$), we find that
$\sim26\%$ of the stars with $-3.2 \le$ [Fe/H] $\le -3.0$ are CEMP
stars, while $\sim39\%$ of the stars below [Fe/H] = $-3.2$ are CEMP
stars. Note also that, below a metallicity of -3.2, the great majority
of the stars shown in Figure 5 are indeed CEMP stars, while the non-CEMP
stars are the dominant fraction above this metallicity. Since over half of our stars have only upper
limits for [Fe/H], we have repeated the exercise with the SSPP estimates
of [Fe/H] for our full sample. For [Fe/H] $\le -3.0$, we obtain a
cumulative frequency of CEMP stars of $\sim 32\%$; for [Fe/H] $\le
-3.5$, the cumulative frequency increases to $\sim42\%$. These results are consistent with the previous calculation within
Poisson errors on the fractions (due to the small sample 
sizes involved, these are on the order of 10\%).

There is no evidence of a correlation between [C/Fe] and [Ca(Mg)/Fe],
nor with the SSPP [$\alpha$/Fe] estimates, as Figure
\ref{cafe_mgfe_cfe_err} shows. Two stars exhibit high [C/Fe] ($> +0.9$)
as well as [Mg/Fe] ($> +0.6$), but normal [Ca/Fe] and [$\alpha$/Fe].


\begin{figure}[!!h]
\center
\includegraphics[scale=0.36, angle=90]{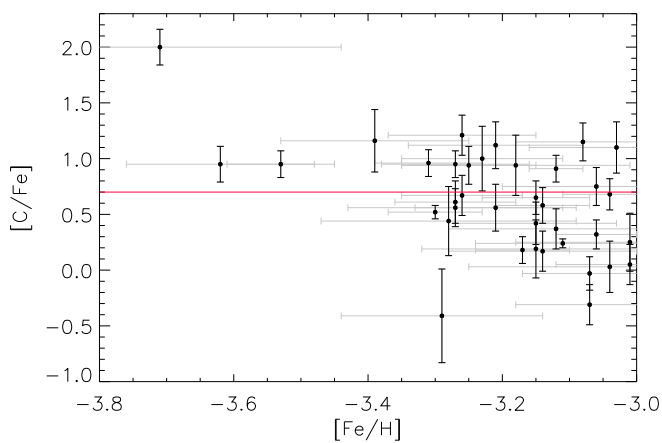} 
\caption{[C/Fe] estimates from the SSPP, as a function of our [Fe/H]
estimate. The red line indicates the level of [C/Fe] $> +0.7$ used to
evaluate whether a star is considered carbon enhanced.}
\label{feh_cfe}
\end{figure}

\begin{figure}[!!h]
\center
\includegraphics[scale=0.5]{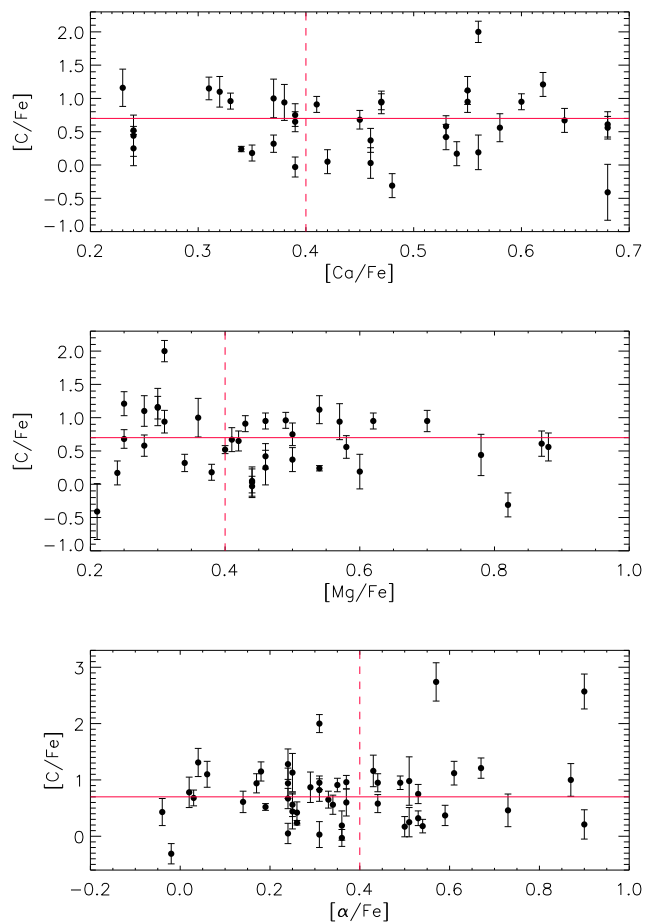} 
\caption{[C/Fe] estimates from the SSPP, as a function of our [Ca/Fe]
and [Mg/Fe] estimates (upper two panels, respectively). The lower panel
shows [C/Fe] as a function of the SSPP [$\alpha$/Fe] estimates. The red
line indicates the level of [C/Fe] $> +0.7$ used to evaluate whether a
star is considered carbon enhanced.}
\label{cafe_mgfe_cfe_err}
\end{figure}

\subsection{SDSS J134144.61$+$474128.6}

We call attention to the CEMP star SDSS J134144.61$+$474128.6 in
our sample, which is a bright ($g = 11.90$) star with [Fe/H]$ = -3.27$
(the SSPP estimate of [Fe/H] = $-2.95$) and [C/Fe] $= +0.95$, identified
during the course of the target search carried out for the MARVELS
sub-survey (see the discussion of the MARVELS pre-survey in Rani et al.
2016). It is of interest that this star also exhibits elevated magnesium
([Mg/Fe] $= +0.62$), which is often found for CEMP-no stars (see, e.g.,
Norris et al. 2013). The absolute carbon abundance, $A(C) = \log\,\epsilon = 6.11$, 
places it on the ``low-C'' band that is associated
with most CEMP-no stars (see, e.g., Bonifacio et al. 2015; Yoon et al.,
in prep.). We note that the other bright EMP star in our sample,
discussed by Rani et al. 2016 (SDSS~J134338.67+484426.6), is not
carbon enhanced. Only $\sim$20 EMP stars have been identified to date
that are as bright as these two stars, hence we plan to obtain a
higher-resolution, higher-S/N spectrum of SDSS J1341$+$4741 in the near
future.

\onllongtab{
\begin{longtable}{lcrrcccccr}
\caption{Equatorial coordinates, SDSS magnitudes and heliocentric velocities for our SDSS/SEGUE and BOSS stellar sample.}\label{SEGUE_coor}\\
\hline
\hline
ID & Plate-MJD-Fiber & $\alpha$ (J2000) & $\delta$ (J2000) & $u$ & $g$ & $r$ & $i$ & $z$ & $v_{hel}$ (km $\rm s^{-1}$) \\
\hline
\endfirsthead
\caption{Continued.} \\
\hline
ID & Plate-MJD-Fiber & $\alpha$ (J2000) & $\delta$ (J2000) & $u$ & $g$ & $r$ & $i$ & $z$ & $v_{hel}$ (km $\rm s^{-1}$) \\
\hline
\endhead
\hline
\endfoot
\hline
\endlastfoot
   SDSS J110827.88$+$613419.0 &  0775-52295-0311  &    167.11614  &     61.57199  &   17.78  &   16.96  &   16.67   &   16.57    &   16.54  &    $-$205.3   \\
   SDSS J115906.16$+$542512.6 &  1018-52672-0268  &    179.77574  &     54.42016  &   17.11  &   16.18  &   15.77   &   15.63    &   15.59  &       182.9   \\
   SDSS J102628.72$+$435742.0 &  1429-52990-0271  &    156.61971  &     43.96169  &   18.04  &   17.21  &   16.90   &   16.81    &   16.74  &        16.7   \\
   SDSS J235718.91$-$005247.7 &  1489-52991-0251  &    359.32881  &   $-$0.87994  &   17.20  &   16.06  &   15.43   &   15.15    &   14.99  &     $-$14.5   \\
   SDSS J191104.68$+$781833.6 &  1857-53182-0438  &    287.76969  &     78.30932  &   18.14  &   17.08  &   16.57   &   16.38    &   16.28  &    $-$233.6   \\
   SDSS J080428.21$+$515303.0 &  1870-53383-0002  &    121.11755  &     51.88419  &   17.44  &   16.48  &   16.09   &   15.94    &   15.85  &    $-$256.1   \\
   SDSS J204735.17$+$001137.9 &  1908-53239-0599  &    311.89656  &     0.19386   &   18.12  &   17.09  &   16.60   &   16.41    &   16.26  &    $-$206.6   \\
   SDSS J115221.93$+$385608.9 &  2027-53433-0324  &    178.09140  &     38.93580  &   17.48  &   16.64  &   16.26   &   16.10    &   16.09  &        36.5   \\
   SDSS J014036.21$+$234458.1 &  2044-53327-0515  &     25.15093  &     23.74947  &   16.78  &   15.82  &   15.35   &   15.12    &   15.03  &    $-$191.0   \\
   SDSS J080057.48$+$070550.5 &  2056-53463-0362  &    120.23952  &      7.09740  &   19.50  &   18.02  &   17.43   &   17.13    &   16.95  &        87.7   \\
   SDSS J161313.51$+$530909.6 &  2176-54243-0614  &    243.30636  &     53.15271  &   17.86  &   16.74  &   16.23   &   16.02    &   15.91  &         1.5   \\
   SDSS J160926.70$+$164743.2 &  2177-54557-0358  &    242.36125  &     16.79535  &   19.16  &   18.02  &   17.41   &   17.13    &   17.01  &     $-$27.8   \\ 
   SDSS J160431.54$+$043220.6 &  2178-54629-0342  &    241.13144  &      4.53908  &   17.74  &   16.72  &   16.23   &   16.03    &   15.94  &     $-$78.0   \\
   SDSS J180516.76$+$231107.7 &  2184-53534-0184  &    271.31988  &     23.18549  &   18.42  &   17.20  &   16.60   &   16.29    &   16.14  &    $-$188.4   \\
   SDSS J160142.38$+$053026.9 &  2186-54327-0277  &    240.42663  &      5.50749  &   19.70  &   18.55  &   18.03   &   17.79    &   17.66  &     $-$31.2   \\
   SDSS J175028.71$+$254434.0 &  2194-53904-0596  &    267.61965  &     25.74279  &   19.68  &   18.68  &   18.10   &   17.83    &   17.71  &    $-$407.3   \\
   SDSS J204524.03$+$150825.3 &  2250-53566-0249  &    311.35016  &     15.14041  &   17.84  &   16.76  &   16.22   &   15.94    &   15.83  &    $-$418.3   \\
   SDSS J220646.20$-$092545.7 &  2309-54441-0290  &    331.69252  &   $-$9.42937  &   16.68  &   15.36  &   14.74   &   14.44    &   14.28  &        13.2   \\
   SDSS J003910.76$+$082821.8 &  2312-53709-0575  &      9.79483  &      8.47273  &   19.09  &   18.01  &   17.44   &   17.19    &   17.05  &    $-$217.8   \\
   SDSS J011323.59$+$002631.5 &  2313-53726-0624  &     18.34832  &      0.44211  &   18.68  &   17.20  &   16.53   &   16.23    &   16.05  &         3.9   \\
   SDSS J225629.81$+$071400.5 &  2325-54082-0397  &    344.12424  &      7.23349  &   20.00  &   18.85  &   18.23   &   18.00    &   17.86  &    $-$179.3   \\
   SDSS J031119.95$+$060850.8 &  2335-53730-0465  &     47.83317  &      6.14746  &   18.21  &   17.17  &   16.59   &   16.32    &   16.13  &        40.5   \\
   SDSS J031357.33$+$050741.2 &  2340-53733-0157  &     48.48890  &      5.12813  &   20.11  &   19.05  &   18.39   &   18.11    &   17.92  &    $-$229.8   \\
   SDSS J023944.44$+$284349.7 &  2442-54065-0500  &     39.93519  &     28.73049  &   19.13  &   17.68  &   16.97   &   16.60    &   16.40  &     $-$17.0   \\
   SDSS J121042.14$+$402645.2 &  2452-54178-0506  &    182.67561  &     40.44589  &   17.99  &   16.96  &   16.53   &   16.31    &   16.22  &     $-$63.1   \\ 
   SDSS J155133.86$+$260423.0 &  2459-54544-0183  &    237.89113  &     26.07307  &   18.55  &   17.54  &   17.04   &   16.88    &   16.77  &     $-$99.5   \\
   SDSS J174235.90$+$643411.2 &  2561-54597-0623  &    265.64959  &     64.56979  &   20.61  &   19.20  &   18.58   &   18.30    &   18.12  &    $-$195.3   \\
   SDSS J215731.90$+$450540.7 &  2566-54333-0198  &    329.38296  &     45.09467  &   19.21  &   18.12  &   17.55   &   17.27    &   17.12  &    $-$322.8   \\
   SDSS J051803.47$+$180544.0 &  2668-54084-0368  &     79.51446  &     18.09557  &   18.35  &   17.10  &   16.46   &   16.16    &   15.98  &      $-$3.6   \\
   SDSS J063055.57$+$255243.7 &  2696-54167-0214  &     97.73157  &     25.87881  &   19.16  &   18.07  &   17.45   &   17.16    &   16.98  &        49.3   \\
   SDSS J173302.14$+$090813.3 &  2797-54616-0565  &    263.25895  &      9.13706  &   18.35  &   17.23  &   16.66   &   16.44    &   16.24  &    $-$246.0   \\
   SDSS J173532.15$+$444635.8 &  2799-54368-0502  &    263.88400  &     44.77663  &   17.22  &   15.97  &   15.38   &   15.10    &   15.00  &    $-$123.4   \\
   SDSS J170339.59$+$283650.1 &  2808-54524-0510  &    255.91500  &     28.61386  &   17.17  &   15.91  &   15.26   &   14.97    &   14.84  &    $-$174.0   \\
   SDSS J170555.89$+$285305.8 &  2808-54524-0543  &    256.48289  &     28.88498  &   18.56  &   17.38  &   16.81   &   16.54    &   16.41  &    $-$211.6   \\
   SDSS J170743.57$+$283643.1 &  2808-54524-0639  &    256.93154  &     28.61200  &   18.63  &   17.65  &   17.13   &   16.91    &   16.83  &    $-$181.4   \\
   SDSS J165835.20$+$272629.5 &  2829-54623-0281  &    254.64671  &     27.44155  &   19.78  &   18.72  &   18.23   &   18.03    &   17.88  &        63.6   \\
   SDSS J085804.66$+$034515.7 &  2888-54529-0566  &    134.51943  &      3.75437  &   19.44  &   18.32  &   17.84   &   17.61    &   17.49  &       251.1   \\
   SDSS J085934.47$+$040232.3 &  2888-54529-0615  &    134.89367  &      4.04233  &   17.26  &   16.12  &   15.62   &   15.39    &   15.29  &       155.2   \\ 
   SDSS J093339.24$+$310245.4 &  2889-54530-0368  &    143.41353  &     31.04594  &   18.91  &   17.60  &   17.07   &   16.81    &   16.67  &       179.5   \\
   SDSS J130615.48$+$390535.3 &  2900-54569-0312  &    196.56452  &     39.09317  &   18.14  &   17.29  &   16.64   &   16.45    &   16.36  &    $-$356.0   \\
   SDSS J144759.68$+$001308.3 &  2909-54653-0496  &    221.99868  &      0.21899  &   18.60  &   17.55  &   17.09   &   16.88    &   16.77  &       285.1   \\
   SDSS J085656.23$+$034410.3 &  2913-54526-0558  &    134.23431  &      3.73621  &   19.36  &   17.76  &   17.05   &   16.71    &   16.51  &       107.4   \\
   SDSS J125330.45$+$191331.1 &  2924-54582-0320  &    193.37690  &     19.22534  &   19.62  &   18.60  &   18.14   &   17.89    &   17.78  &        53.1   \\
   SDSS J143914.31$+$212259.4 &  2964-54632-0489  &    219.80968  &     21.38317  &   17.32  &   16.01  &   15.44   &   15.18    &   15.00  &    $-$198.8   \\
   SDSS J125050.82$+$102520.2 &  2965-54594-0461  &    192.71179  &     10.42229  &   18.55  &   17.21  &   16.63   &   16.36    &   16.22  &        73.6   \\
   SDSS J134144.61$+$474128.6 &  3003-54845-0181  &    205.43579  &     47.69100  &   12.66  &   11.90  &   11.34   &   11.19    &   11.25  &    $-$198.4   \\
   SDSS J134338.67$+$484426.6 &  3003-54845-0453  &    205.91109  &     48.74069  &   13.31  &   12.45  &   12.05   &   11.92    &   11.94  &     $-$97.4   \\
   SDSS J003529.36$-$111405.2 &  3105-54825-0010  &      8.87234  &  $-$11.23480  &   18.50  &   17.18  &   16.55   &   16.26    &   16.11  &    $-$235.4   \\
   SDSS J005420.85$-$003905.4 &  3112-54802-0282  &     13.58690  &   $-$0.65151  &   19.11  &   18.10  &   17.64   &   17.43    &   17.34  &    $-$184.3   \\
   SDSS J035622.42$+$114705.4 &  3121-54749-0496  &     59.09343  &     11.78486  &   19.23  &   18.04  &   17.43   &   17.14    &   16.97  &     $-$19.1   \\
   SDSS J021958.24$-$084955.8 &  3122-54821-0178  &     34.99267  &   $-$8.83218  &   17.42  &   16.42  &   15.97   &   15.72    &   15.68  &     $-$21.5   \\ 
   SDSS J022654.58$-$083951.6 &  3122-54821-0621  &     36.72738  &   $-$8.66437  &   18.89  &   17.85  &   17.42   &   17.19    &   17.11  &     $-$99.8   \\
   SDSS J234403.47$+$151744.6 &  3130-54740-0489  &    356.01446  &     15.29574  &   18.15  &   17.08  &   16.59   &   16.36    &   16.24  &    $-$293.9   \\
   SDSS J093231.65$+$085037.1 &  3151-54804-0328  &    143.13189  &      8.84365  &   18.99  &   17.70  &   17.05   &   16.81    &   16.63  &       173.0   \\
   SDSS J102328.72$+$352919.0 &  3152-54801-0068  &    155.86967  &     35.48863  &   18.99  &   17.98  &   17.57   &   17.38    &   17.33  &     $-$71.1   \\
   SDSS J100537.71$+$022734.1 &  3154-54821-0623  &    151.40716  &      2.45948  &   19.02  &   18.14  &   17.77   &   17.65    &   17.60  &      $-$0.2   \\
   SDSS J112805.85$+$204651.2 &  3170-54859-0571  &    172.02440  &     20.78090  &   17.18  &   16.07  &   15.56   &   15.38    &   15.20  &       217.1   \\
   SDSS J110445.18$+$670752.0 &  3171-54862-0412  &    166.18828  &     67.13112  &   17.05  &   16.00  &   15.52   &   15.33    &   15.21  &      $-$1.4   \\
   SDSS J101600.42$+$172901.1 &  3178-54848-0086  &    154.00179  &     17.48366  &   17.59  &   16.59  &   16.09   &   15.85    &   15.76  &       121.0   \\
   SDSS J120858.66$+$302312.9 &  3180-54864-0372  &    182.24443  &     30.38693  &   17.83  &   16.81  &   16.33   &   16.20    &   16.01  &         4.2   \\
   SDSS J031745.82$+$002304.1 &  3183-54833-0490  &     49.44092  &      0.38450  &   17.79  &   16.80  &   16.35   &   16.14    &   16.02  &       108.3   \\
   SDSS J025938.05$+$010735.3 &  3184-54850-0442  &     44.90854  &      1.12656  &   19.03  &   18.02  &   17.51   &   17.28    &   17.17  &    $-$235.2   \\ 
   SDSS J025956.49$+$005712.7 &  3184-54850-0451  &     44.98521  &      0.95370  &   18.47  &   16.75  &   15.87   &   15.48    &   15.26  &        36.4   \\
   SDSS J031259.10$-$061957.1 &  3186-54833-0328  &     48.24627  &   $-$6.33254  &   18.03  &   17.05  &   16.56   &   16.36    &   16.25  &       228.2   \\
   SDSS J033146.90$+$182530.7 &  3187-54821-0618  &     52.94545  &     18.42520  &   19.02  &   17.58  &   16.87   &   16.53    &   16.33  &     $-$92.0   \\
   SDSS J100401.24$+$420150.4 &  3194-54833-0090  &    151.00520  &     42.03068  &   18.25  &   17.38  &   17.04   &   16.91    &   16.85  &        42.5   \\
   SDSS J083235.90$+$134538.3 &  3195-54832-0153  &    128.14960  &     13.76065  &   19.31  &   18.06  &   17.40   &   17.12    &   16.94  &       231.2   \\
   SDSS J045637.24$-$044123.0 &  3209-54906-0287  &     74.15517  &   $-$4.68971  &   19.01  &   17.86  &   17.35   &   17.14    &   16.98  &        90.7   \\
   SDSS J120441.38$+$120111.5 &  3214-54866-0429  &    181.17244  &     12.01987  &   17.35  &   16.43  &   16.09   &   15.94    &   15.86  &        78.4   \\
   SDSS J082511.45$+$163459.9 &  3230-54860-0367  &    126.29774  &     16.58332  &   18.61  &   17.59  &   17.08   &   16.86    &   16.78  &        23.0   \\
   SDSS J112827.62$-$002232.1 &  3233-54891-0490  &    172.11511  &   $-$0.37560  &   17.05  &   15.78  &   15.19   &   14.91    &   14.75  &       175.1   \\
   SDSS J022534.63$+$233504.5 &  3241-54884-0233  &     36.39428  &     23.58458  &   18.96  &   17.97  &   17.51   &   17.33    &   17.21  &    $-$193.2   \\
   SDSS J115019.10$+$135535.4 &  3245-54894-0609  &    177.57961  &     13.92654  &   18.65  &   17.29  &   16.67   &   16.37    &   16.21  &       137.7   \\
   SDSS J104702.87$+$214308.4 &  3251-54882-0208  &    161.76198  &     21.71902  &   18.58  &   17.53  &   17.07   &   16.85    &   16.75  &     $-$39.1   \\ 
   SDSS J123404.57$+$134411.4 &  3254-54889-0128  &    188.51907  &     13.73652  &   17.80  &   16.87  &   16.46   &   16.29    &   16.19  &    $-$102.1   \\
   SDSS J104631.66$+$283819.7 &  3263-54887-0350  &    161.63196  &     28.63881  &   18.62  &   17.60  &   17.13   &   16.88    &   16.71  &       134.2   \\
   SDSS J091844.56$+$431219.3 &  3264-54889-0214  &    139.68568  &     43.20540  &   19.43  &   18.21  &   17.60   &   17.35    &   17.17  &       157.5   \\
   SDSS J082452.06$+$645145.6 &  3286-54910-0003  &    126.21685  &     64.86279  &   19.24  &   18.37  &   18.01   &   17.84    &   17.79  &    $-$141.5   \\
   SDSS J104617.20$+$172712.3 &  3299-54908-0450  &    161.57169  &     17.45342  &   17.95  &   16.38  &   15.65   &   15.33    &   15.17  &       170.0   \\
   SDSS J132250.59$+$012342.9 &  3307-54970-0529  &    200.71082  &      1.39528  &   17.47  &   16.32  &   15.82   &   15.60    &   15.49  &       105.5   \\
   SDSS J152853.64$+$065510.2 &  3308-54919-0513  &    232.22353  &      6.91952  &   19.10  &   17.61  &   16.97   &   16.64    &   16.47  &     $-$20.9   \\
   SDSS J092540.08$+$094104.3 &  3319-54915-0230  &    141.41699  &      9.68455  &   18.00  &   16.63  &   15.99   &   15.67    &   15.49  &       245.6   \\
   SDSS J111104.14$+$413942.6 &  3326-54943-0487  &    167.76728  &     41.66185  &   18.04  &   17.07  &   16.64   &   16.46    &   16.35  &     $-$91.4   \\
   SDSS J115525.43$+$474420.1 &  3331-54977-0355  &    178.85600  &     47.73893  &   18.72  &   17.83  &   17.44   &   17.28    &   17.25  &        67.6   \\
   SDSS J132146.89$+$322733.2 &  3377-54950-0119  &    200.44541  &     32.45924  &   19.46  &   18.49  &   18.01   &   17.87    &   17.75  &        60.0   \\
   SDSS J142928.29$+$295542.7 &  3384-54948-0172  &    217.36790  &     29.92855  &   17.58  &   16.36  &   15.80   &   15.52    &   15.41  &    $-$216.5   \\
   SDSS J144726.06$+$224552.7 &  3387-54951-0261  &    221.85861  &     22.76466  &   19.12  &   18.11  &   17.59   &   17.41    &   17.25  &    $-$250.6   \\ 
   SDSS J145056.30$+$234116.4 &  3387-54951-0430  &    222.73460  &     23.68791  &   18.36  &   17.23  &   16.71   &   16.49    &   16.35  &       136.1   \\
   SDSS J145124.04$+$093845.2 &  3388-54947-0187  &    222.85019  &      9.64589  &   18.97  &   17.89  &   17.45   &   17.25    &   17.14  &       115.1   \\
   SDSS J144442.53$+$214647.0 &  3407-54971-0353  &    221.17724  &     21.77972  &   18.73  &   17.70  &   17.22   &   17.00    &   16.86  &     $-$57.4   \\
   SDSS J165618.30$+$342523.0 &  3457-54984-0589  &    254.07628  &     34.42308  &   17.20  &   15.68  &   15.06   &   14.78    &   14.60  &    $-$391.6   \\
   SDSS J013643.25$+$010525.4 &  3639-55146-0544  &     24.18024  &      1.09041  &   18.65  &   17.77  &   17.39   &   17.25    &   17.18  &     $-$66.8   \\
   SDSS J023248.01$+$003735.3 &  3647-55827-0970  &     38.20007  &      0.62648  &   18.61  &   17.77  &   17.46   &   17.36    &   17.33  &       185.5   \\
   SDSS J024458.86$+$011823.8 &  3651-55247-0784  &     41.24527  &      1.30664  &   19.05  &   18.15  &   17.78   &   17.64    &   17.56  &     $-$72.3   \\
   SDSS J074449.38$+$471237.3 &  3665-55247-0074  &    116.20575  &     47.21045  &   17.63  &   16.72  &   16.36   &   16.19    &   16.13  &       231.8   \\
   SDSS J141249.08$+$013206.7 &  4030-55634-0390  &    213.20450  &      1.53520  &   18.94  &   18.08  &   17.69   &   17.53    &   17.45  &       141.9   \\
   SDSS J135331.01$-$032930.2 &  4041-55361-0401  &    208.37924  &   $-$3.49170  &   17.72  &   16.79  &   16.51   &   16.41    &   16.35  &       156.3   \\
   SDSS J134916.87$+$002622.7 &  4043-55630-0998  &    207.32032  &      0.43966  &   19.03  &   18.21  &   17.83   &   17.70    &   17.64  &     $-$36.0   \\
   SDSS J213242.72$+$040533.9 &  4084-55447-0163  &    323.17803  &      4.09275  &   18.70  &   17.88  &   17.56   &   17.45    &   17.38  &    $-$108.0   \\ 
   SDSS J084214.91$+$195254.9 &  5176-56221-0514  &    130.56215  &     19.88193  &   17.21  &   16.37  &   16.04   &   15.88    &   15.86  &       102.3   \\
   SDSS J013257.20$+$252558.5 &  5695-55978-0479  &     23.23836  &     25.43293  &   18.84  &   17.92  &   17.56   &   17.42    &   17.38  &    $-$232.7   \\
   SDSS J091308.43$+$595702.4 &  5712-56602-0618  &    138.28509  &     59.95079  &   17.23  &   16.31  &   15.92   &   15.76    &   15.74  &    $-$392.1   \\
   SDSS J092931.26$+$592711.7 &  5716-56684-0478  &    142.38015  &     59.45335  &   17.84  &   16.96  &   16.65   &   16.53    &   16.52  &        95.7   \\
   SDSS J092147.33$+$261936.6 &  5797-56273-0288  &    140.44724  &     26.32684  &   18.56  &   17.71  &   17.37   &   17.25    &   17.22  &        86.9   \\
   SDSS J004705.27$+$233956.0 &  6287-56221-0822  &     11.77200  &     23.66557  &   19.07  &   18.26  &   17.90   &   17.77    &   17.70  &     $-$65.4   \\
   SDSS J231319.28$+$211528.5 &  6592-56535-0046  &    348.33036  &     21.25794  &   20.52  &   19.31  &   18.72   &   18.43    &   18.28  &    $-$136.5   \\
   SDSS J014451.69$+$333127.7 &  6604-56337-0611  &     26.21538  &     33.52438  &   16.96  &   16.13  &   15.75   &   15.60    &   15.55  &    $-$143.4   \\
\hline\hline
\end{longtable}
}

\onllongtab{
\begin{longtable}{lccccccc}
\caption{Stellar parameters $T_{\rm eff}$, $\log g$ and [M/H] from our analysis and the SSPP for SDSS/SEGUE and BOSS EMP stars.}\label{SEGUE_sp}\\
\hline
\hline
Plate-MJD-Fiber & $T_{\rm eff}$ (K) & $\log g$ & [M/H] & $\sigma$[M/H] & $T_{\rm eff}$ SSPP (K) & $\log g$ SSPP & [Fe/H] SSPP \\
\hline
\endfirsthead
\caption{Continued.} \\
\hline
Plate-MJD-Fiber & $T_{\rm eff}$ (K) & $\log g$ & [M/H] & $\sigma$[M/H] & $T_{\rm eff}$ SSPP (K) & $\log g$ SSPP & [Fe/H] SSPP \\
\hline
\endhead
\hline
\endfoot
\hline
\endlastfoot
   0775-52295-0311     &    5512      &      0.51      &     $-3.17$      &      0.91      &      6202  &  4.20  &  $-2.52$   \\
   1018-52672-0268     &    5488      &      1.62      &     $-3.24$      &      0.03      &      5852  &  3.34  &  $-3.02$   \\
   1429-52990-0271     &    5520      &      0.53      &     $-3.29$      &      0.31      &      6003  &  3.36  &  $-2.94$   \\
   1489-52991-0251     &    5000      &      4.03      &     $-3.32$      &      0.02      &      5277  &  3.58  &  $-3.79$   \\
   1857-53182-0438     &    5309      &      1.12      &     $-3.17$      &      0.04      &      5641  &  3.75  &  $-2.84$   \\
   1870-53383-0002     &    5529      &      1.17      &     $-$3.24      &      0.04      &      5989  &  3.55  &  $-$2.85   \\
   1908-53239-0599     &    5498      &      2.79      &     $-$3.03      &      0.04      &      5700  &  3.66  &  $-$3.06   \\
   2027-53433-0324     &    5865      &      4.49      &     $-$3.04      &      0.07      &      5911  &  4.05  &  $-$3.34   \\
   2044-53327-0515     &    6064      &      4.48      &     $-$3.42      &      0.05      &      6172  &  3.66  &  $-$3.30   \\
   2056-53463-0362     &    4781      &      2.23      &     $-$3.20      &      0.11      &      5349  &  3.56  &  $-$3.09   \\
   2176-54243-0614     &    4751      &      3.15      &     $-$3.02      &      0.07      &      5444  &  2.77  &  $-$2.83   \\
   2177-54557-0358     &    5065      &      4.49      &     $-$3.15      &      0.07      &      5191  &  3.29  &  $-$3.34   \\
   2178-54629-0342     &    5597      &      4.23      &     $-$3.00      &      0.11      &      5628  &  2.91  &  $-$3.11   \\
   2184-53534-0184     &    5116      &      0.63      &     $-$3.02      &      0.08      &      5440  &  2.82  &  $-$2.80   \\
   2186-54327-0277     &    5196      &      0.52      &     $-$3.25      &      0.53      &      5583  &  2.77  &  $-$3.23   \\
   2194-53904-0596     &    5327      &      4.49      &     $-$3.23      &      0.16      &      5397  &  3.86  &  $-$3.31   \\ 
   2250-53566-0249     &    5035      &      2.23      &     $-$3.59      &      0.05      &      5422  &  2.97  &  $-$3.38   \\
   2309-54441-0290     &    4902      &      1.49      &     $-$3.08      &      0.02      &      5175  &  2.43  &  $-$3.02   \\
   2312-53709-0575     &    5110      &      0.52      &     $-$3.63      &      0.51      &      5326  &  2.33  &  $-$3.55   \\
   2313-53726-0624     &    4770      &      0.51      &     $-$3.05      &      0.32      &      4928  &  1.62  &  $-$3.00   \\
   2325-54082-0397     &    5086      &      1.69      &     $-$3.08      &      0.06      &      5246  &  1.92  &  $-$3.16   \\
   2335-53730-0465     &    5255      &      0.55      &     $-$3.04      &      0.16      &      5929  &  4.03  &  $-$2.49   \\
   2340-53733-0157     &    5026      &      0.52      &     $-$3.16      &      0.75      &      5519  &  3.58  &  $-$2.79   \\
   2442-54065-0500     &    4752      &      2.37      &     $-$3.09      &      0.27      &      5412  &  2.09  &  $-$2.92   \\
   2452-54178-0506     &    5383      &      3.06      &     $-$3.07      &      0.04      &      5638  &  3.31  &  $-$3.09   \\
   2459-54544-0183     &    5653      &      3.89      &     $-$3.44      &      0.05      &      5697  &  3.56  &  $-$3.66   \\
   2561-54597-0623     &    4846      &      4.46      &     $-$3.58      &      0.17      &      4518  &  3.74  &  $-$3.52   \\
   2566-54333-0198     &    5257      &      0.51      &     $-$3.12      &      0.59      &      6005  &  4.23  &  $-$2.27   \\
   2668-54084-0368     &    5564      &      0.51      &     $-$3.00      &      0.27      &      6476  &  4.13  &  $-$2.31   \\
   2696-54167-0214     &    5399      &      0.53      &     $-$3.77      &      0.18      &      6088  &  2.87  &  $-$3.30   \\
   2797-54616-0565     &    5306      &      0.77      &     $-$3.23      &      0.03      &      5700  &  3.62  &  $-$2.85   \\
   2799-54368-0502     &    5118      &      1.27      &     $-$3.08      &      0.02      &      5358  &  3.19  &  $-$3.06   \\
   2808-54524-0510     &    4917      &      4.24      &     $-$3.40      &      0.04      &      5300  &  4.57  &  $-$3.44   \\
   2808-54524-0543     &    5136      &      2.42      &     $-$3.10      &      0.03      &      5438  &  2.59  &  $-$3.01   \\
   2808-54524-0639     &    5614      &      4.49      &     $-$3.08      &      0.09      &      5687  &  3.23  &  $-$2.95   \\
   2829-54623-0281     &    5274      &      0.63      &     $-$3.45      &      0.09      &      5441  &  2.75  &  $-$3.27   \\
   2888-54529-0566     &    5250      &      1.00      &     $-$3.07      &      0.04      &      5540  &  3.24  &  $-$2.87   \\
   2888-54529-0615     &    5194      &      0.95      &     $-$3.13      &      0.01      &      5520  &  3.18  &  $-$3.05   \\
   2889-54530-0368     &    5160      &      1.28      &     $-$3.01      &      0.03      &      5274  &  2.08  &  $-$2.96   \\
   2900-54569-0312     &    5462      &      4.49      &     $-$3.09      &      0.10      &      5505  &  4.40  &  $-$3.50   \\
   2909-54653-0496     &    5599      &      4.27      &     $-$3.25      &      0.10      &      5634  &  3.01  &  $-$3.26   \\
   2913-54526-0558     &    4873      &      0.82      &     $-$3.04      &      0.02      &      4812  &  0.83  &  $-$3.46   \\
   2924-54582-0320     &    5288      &      3.79      &     $-$3.04      &      0.06      &      5390  &  2.57  &  $-$2.94   \\
   2964-54632-0489     &    5113      &      1.74      &     $-$3.04      &      0.01      &      5353  &  2.94  &  $-$3.10   \\
   2965-54594-0461     &    5111      &      2.94      &     $-$3.08      &      0.02      &      5361  &  2.92  &  $-$3.28   \\
   3003-54845-0181     &    5254      &      0.51      &     $-$3.24      &      0.36      &      5808  &  3.38  &  $-$2.95   \\
   3003-54845-0453     &    5307      &      0.51      &     $-$3.88      &      0.50      &      6006  &  3.39  &  $-$3.42   \\
   3105-54825-0010     &    4935      &      0.72      &     $-$3.06      &      0.04      &      5364  &  2.91  &  $-$2.88   \\
   3112-54802-0282     &    5286      &      0.62      &     $-$3.18      &      0.07      &      5590  &  3.27  &  $-$2.83   \\
   3121-54749-0496     &    5305      &      0.85      &     $-$3.03      &      0.02      &      6002  &  4.08  &  $-$2.48   \\
   3122-54821-0178     &    5689      &      3.94      &     $-$3.34      &      0.03      &      5608  &  3.08  &  $-$3.70   \\
   3122-54821-0621     &    5227      &      0.61      &     $-$3.09      &      0.07      &      5623  &  3.36  &  $-$2.85   \\
   3130-54740-0489     &    5238      &      0.96      &     $-$3.38      &      0.01      &      5548  &  3.03  &  $-$3.32   \\
   3151-54804-0328     &    5004      &      1.20      &     $-$3.07      &      0.03      &      5360  &  2.58  &  $-$3.19   \\
   3152-54801-0068     &    5223      &      2.28      &     $-$3.50      &      0.04      &      5527  &  2.45  &  $-$3.67   \\
   3154-54821-0623     &    5428      &      0.51      &     $-$3.48      &      0.51      &      5989  &  3.67  &  $-$3.10   \\
   3170-54859-0571     &    5308      &      3.24      &     $-$3.05      &      0.02      &      5516  &  3.13  &  $-$3.22   \\
   3171-54862-0412     &    5283      &      1.21      &     $-$3.06      &      0.01      &      5569  &  3.26  &  $-$3.02   \\
   3178-54848-0086     &    5390      &      4.49      &     $-$3.38      &      0.08      &      5473  &  3.72  &  $-$3.23   \\
   3180-54864-0372     &    5498      &      4.05      &     $-$3.45      &      0.03      &      5513  &  3.02  &  $-$3.66   \\
   3183-54833-0490     &    5780      &      3.72      &     $-$3.23      &      0.02      &      5879  &  3.36  &  $-$3.36   \\
   3184-54850-0442     &    5423      &      4.49      &     $-$3.21      &      0.09      &      5543  &  3.73  &  $-$3.15   \\
   3184-54850-0451     &    4750      &      4.49      &     $-$3.19      &      0.21      &      4686  &  4.71  &  $-$3.44   \\
   3186-54833-0328     &    5130      &      0.51      &     $-$3.15      &      0.52      &      5682  &  3.83  &  $-$2.88   \\
   3187-54821-0618     &    4830      &      2.60      &     $-$3.05      &      0.03      &      5312  &  2.69  &  $-$2.93   \\
   3194-54833-0090     &    5840      &      3.05      &     $-$3.16      &      0.05      &      5934  &  3.02  &  $-$3.41   \\
   3195-54832-0153     &    4954      &      2.55      &     $-$3.19      &      0.06      &      5199  &  2.31  &  $-$3.10   \\
   3209-54906-0287     &    5060      &      0.81      &     $-$3.09      &      0.04      &      5469  &  3.05  &  $-$2.74   \\
   3214-54866-0429     &    5873      &      3.94      &     $-$3.51      &      0.03      &      5917  &  3.04  &  $-$3.67   \\
   3230-54860-0367     &    5524      &      3.68      &     $-$3.28      &      0.03      &      5637  &  3.39  &  $-$3.32   \\
   3233-54891-0490     &    5098      &      2.44      &     $-$3.32      &      0.02      &      5474  &  3.50  &  $-$3.32   \\
   3241-54884-0233     &    5492      &      0.54      &     $-$3.15      &      0.18      &      6152  &  3.83  &  $-$2.51   \\
   3245-54894-0609     &    4866      &      0.71      &     $-$3.19      &      0.03      &      5267  &  2.63  &  $-$3.06   \\
   3251-54882-0208     &    5282      &      3.57      &     $-$3.10      &      0.05      &      5469  &  3.43  &  $-$3.17   \\
   3254-54889-0128     &    5872      &      3.90      &     $-$3.53      &      0.05      &      5980  &  3.71  &  $-$3.46   \\
   3263-54887-0350     &    5207      &      0.72      &     $-$3.33      &      0.04      &      5539  &  2.62  &  $-$3.15   \\
   3264-54889-0214     &    5018      &      2.03      &     $-$3.35      &      0.06      &      5196  &  2.21  &  $-$3.88   \\
   3286-54910-0003     &    5516      &      0.52      &     $-$3.43      &      0.53      &      6052  &  3.55  &  $-$3.02   \\
   3299-54908-0450     &    4881      &      0.51      &     $-$3.04      &      0.12      &      4855  &  1.97  &  $-$3.23   \\
   3307-54970-0529     &    5234      &      0.84      &     $-$3.60      &      0.02      &      5536  &  3.25  &  $-$3.68   \\
   3308-54919-0513     &    4915      &      0.96      &     $-$3.02      &      0.03      &      5190  &  2.22  &  $-$3.02   \\
   3319-54915-0230     &    5114      &      1.77      &     $-$3.26      &      0.02      &      5126  &  2.51  &  $-$3.40   \\
   3326-54943-0487     &    5694      &      4.06      &     $-$3.02      &      0.03      &      5720  &  3.61  &  $-$3.20   \\
   3331-54977-0355     &    5760      &      4.49      &     $-$3.23      &      0.10      &      5879  &  3.82  &  $-$3.10   \\
   3377-54950-0119     &    5220      &      3.33      &     $-$3.29      &      0.08      &      5468  &  3.39  &  $-$3.23   \\
   3384-54948-0172     &    5133      &      1.96      &     $-$3.21      &      0.02      &      5345  &  2.61  &  $-$3.25   \\
   3387-54951-0261     &    5299      &      4.49      &     $-$3.37      &      0.14      &      5520  &  3.84  &  $-$3.35   \\
   3387-54951-0430     &    5231      &      3.12      &     $-$3.04      &      0.03      &      5457  &  3.31  &  $-$2.94   \\
   3388-54947-0187     &    5513      &      3.73      &     $-$3.28      &      0.06      &      5606  &  3.42  &  $-$3.35   \\
   3407-54971-0353     &    5255      &      2.69      &     $-$3.04      &      0.04      &      5481  &  2.70  &  $-$3.08   \\
   3457-54984-0589     &    5009      &      0.99      &     $-$3.11      &      0.01      &      5352  &  2.84  &  $-$2.99   \\
   3639-55146-0544     &    5488      &      0.52      &     $-$3.39      &      0.37      &      6015  &  3.29  &  $-$2.96   \\
   3647-55827-0970     &    5375      &      0.51      &     $-$3.16      &      0.52      &      6184  &  3.68  &  $-$2.30   \\
   3651-55247-0784     &    5322      &      0.51      &     $-$3.11      &      0.52      &      5970  &  3.44  &  $-$2.50   \\
   3665-55247-0074     &    5503      &      0.51      &     $-$3.37      &      0.27      &      6164  &  3.78  &  $-$2.70   \\
   4030-55634-0390     &    5572      &      1.02      &     $-$3.69      &      0.06      &      5993  &  3.09  &  $-$3.42   \\
   4041-55361-0401     &    5595      &      0.52      &     $-$3.89      &      0.17      &      6297  &  3.21  &  $-$3.42   \\
   4043-55630-0998     &    4750      &      3.23      &     $-$3.09      &      0.27      &      5999  &  3.37  &  $-$3.60   \\
   4084-55447-0163     &    5268      &      0.51      &     $-$3.53      &      0.70      &      6258  &  3.52  &  $-$2.60   \\ 
   5176-56221-0514     &    5426      &      0.51      &     $-$3.33      &      0.39      &      6112  &  3.86  &  $-$2.86   \\
   5695-55978-0479     &    5362      &      0.50      &     $-$3.36      &      1.27      &      6293  &  4.27  &  $-$2.32   \\
   5712-56602-0618     &    5292      &      0.50      &     $-$3.05      &      0.19      &      6067  &  4.02  &  $-$2.52   \\
   5716-56684-0478     &    5419      &      0.54      &     $-$3.40      &      0.16      &      6216  &  3.83  &  $-$2.56   \\
   5797-56273-0288     &    5419      &      0.51      &     $-$3.53      &      0.35      &      6116  &  3.59  &  $-$3.13   \\
   6287-56221-0822     &    5287      &      0.51      &     $-$3.37      &      0.75      &      5963  &  3.27  &  $-$3.06   \\
   6592-56535-0046     &    5252      &      0.50      &     $-$3.11      &      0.88      &      6122  &  3.31  &  $-$2.51   \\
   6604-56337-0611     &    5375      &      0.50      &     $-$3.20      &      0.46      &      6003  &  3.77  &  $-$2.76   \\
\hline\hline
\end{longtable}
}

\onllongtab{
\begin{longtable}{lrrrrrrrr}
\caption{Chemical abundances for the SDSS/SEGUE and BOSS EMP stars.}\label{SEGUE_ab}\\
\hline
\hline
 Plate-MJD-Fiber & [Fe/H] &  $\sigma$[Fe/H] & [Ca/H] & $\sigma$[Ca/H] & [Mg/H]  & $\sigma$[Mg/H] & [$\alpha$/Fe] SSPP & [C/Fe] SSPP \\
\hline
\endfirsthead
\caption{Continued.} \\
\hline
Plate-MJD-Fiber & [Fe/H] &  $\sigma$[Fe/H] & [Ca/H] & $\sigma$[Ca/H] & [Mg/H]  & $\sigma$[Mg/H] & [$\alpha$/Fe] SSPP & [C/Fe] SSPP \\
\hline
\endhead
\hline
\endfoot
\hline
\endlastfoot
   0775-52295-0311  &     $<-3.00$   &     -    &   $-2.72$     &     0.05  &    $ -2.42$    &      0.26  &     0.31  &     -    \\
   1018-52672-0268  &     $ -3.28$   &  0.22    &   $-2.80$     &     0.02  &    $ -3.14$    &      0.24  &   $-0.30$ &     -    \\
   1429-52990-0271  &     $<-3.00$   &     -    &   $-2.82$     &     0.04  &    $ -3.17$    &      0.28  &       -   &     -    \\
   1489-52991-0251  &     $ -3.62$   &  0.14    &   $-3.07$     &     0.02  &    $ -2.92$    &      0.11  &    0.44   &   0.95   \\
   1857-53182-0438  &     $<-3.20$   &     -    &   $-2.71$     &     0.05  &    $ -2.65$    &      0.37  &    0.30   &     -    \\
   1870-53383-0002  &     $<-3.05$   &     -    &   $-2.83$     &     0.04  &    $ -2.67$    &      0.27  &    0.29   &     -    \\
   1908-53239-0599  &     $<-3.10$   &     -    &   $-2.56$     &     0.03  &    $ -2.89$    &      0.30  &    0.73   &   0.46   \\
   2027-53433-0324  &     $<-3.10$   &     -    &   $-2.63$     &     0.03  &    $ -2.43$    &      0.11  &    0.24   &     -    \\
   2044-53327-0515  &     $<-3.00$   &     -    &   $-3.03$     &     0.03  &    $ -3.01$    &      0.15  &      -    &   2.19   \\
   2056-53463-0362  &     $<-3.20$   &     -    &   $-2.81$     &     0.07  &    $ -2.77$    &      0.30  &    0.51   &   0.98   \\
   2176-54243-0614  &     $ -3.71$   &  0.27    &   $-3.15$     &     0.02  &    $ -3.40$    &      0.10  &    0.31   &   2.00   \\
   2177-54557-0358  &     $ -3.17$   &  0.07    &   $-2.82$     &     0.03  &    $ -2.79$    &      0.07  &    0.54   &   0.18   \\
   2178-54629-0342  &     $<-3.00$   &     -    &   $-2.58$     &     0.02  &    $ -3.16$    &      0.50  &    0.51   &     -    \\
   2184-53534-0184  &     $ -3.04$   &  0.21    &   $-2.58$     &     0.04  &    $ -2.60$    &      0.24  &    0.31   &   0.03   \\
   2186-54327-0277  &     $<-3.07$   &     -    &   $-2.85$     &     0.06  &    $<-3.32$    &       -    &      -    &     -    \\
   2194-53904-0596  &     $<-3.00$   &     -    &   $-3.09$     &     0.07  &    $ -2.21$    &      0.08  &    0.50   &     -    \\ 
   2250-53566-0249  &     $<-3.40$   &     -    &   $-3.22$     &     0.03  &    $ -3.22$    &      0.15  &   $-0.04$ &   0.43   \\
   2309-54441-0290  &     $ -3.06$   &  0.14    &   $-2.67$     &     0.01  &    $ -2.56$    &      0.10  &    0.53   &   0.75   \\
   2312-53709-0575  &     $<-3.20$   &     -    &   $-3.08$     &     0.06  &    $<-3.40$    &       -    &      -    &     -    \\
   2313-53726-0624  &     $ -3.21$   &  0.12    &   $-2.54$     &     0.04  &    $ -2.75$    &      0.32  &    0.42   &     -    \\
   2325-54082-0397  &     $<-3.00$   &     -    &   $-2.70$     &     0.08  &    $<-3.30$    &       -    &      -    &   0.25   \\
   2335-53730-0465  &     $ -3.15$   &  0.17    &   $-2.59$     &     0.02  &    $ -2.55$    &      0.20  &    0.36   &   0.19   \\
   2340-53733-0157  &     $<-3.00$   &     -    &   $-2.56$     &     0.09  &    $ -2.46$    &      0.52  &    0.38   &     -    \\
   2442-54065-0500  &     $<-3.24$   &     -    &   $-2.92$     &     0.05  &    $ -2.94$    &      0.25  &    0.24   &   1.28   \\
   2452-54178-0506  &     $<-3.10$   &     -    &   $-2.66$     &     0.02  &    $ -2.88$    &      0.19  &    0.29   &   0.87   \\
   2459-54544-0183  &     $<-3.10$   &     -    &   $-3.00$     &     0.04  &    $ -2.72$    &      0.15  &    0.11   &     -    \\
   2561-54597-0623  &     $<-3.20$   &     -    &   $-3.28$     &     0.08  &    $ -3.19$    &      0.14  &      -    &     -    \\
   2566-54333-0198  &     $<-3.22$   &     -    &   $-2.61$     &     0.06  &    $ -2.71$    &      0.24  &    0.25   &   1.13   \\
   2668-54084-0368  &     $<-3.00$   &     -    &   $-2.56$     &     0.03  &    $ -2.19$    &      0.33  &    0.55   &     -    \\
   2696-54167-0214  &     $<-3.30$   &     -    &   $-3.33$     &     0.06  &    $ -3.31$    &      0.23  &      -    &     -    \\
   2797-54616-0565  &     $ -3.18$   &  0.17    &   $-2.80$     &     0.02  &    $ -2.61$    &      0.14  &    0.24   &   0.94   \\
   2799-54368-0502  &     $ -3.29$   &  0.15    &   $-2.61$     &     0.01  &    $ -3.08$    &      0.14  &      -    &  $-0.41$ \\
   2808-54524-0510  &     $ -3.30$   &  0.07    &   $-3.06$     &     0.01  &    $ -2.90$    &      0.04  &    0.19   &   0.52   \\
   2808-54524-0543  &     $<-3.46$   &     -    &   $-2.69$     &     0.03  &    $ -2.62$    &      0.14  &    0.04   &   1.31   \\
   2808-54524-0639  &     $<-3.20$   &     -    &   $-2.58$     &     0.04  &    $ -3.01$    &      0.13  &    0.90   &     -    \\
   2829-54623-0281  &     $<-3.10$   &     -    &   $-3.02$     &     0.05  &    $ -2.95$    &      0.35  &      -    &     -    \\
   2888-54529-0566  &     $<-3.30$   &     -    &   $-2.59$     &     0.05  &    $ -2.98$    &      0.26  &      -    &   0.28   \\
   2888-54529-0615  &     $ -3.06$   &  0.06    &   $-2.69$     &     0.01  &    $ -2.72$    &      0.11  &    0.53   &   0.32   \\
   2889-54530-0368  &     $ -3.09$   &  0.15    &   $-2.56$     &     0.03  &    $ -2.64$    &      0.23  &    0.40   &  $-0.41$ \\
   2900-54569-0312  &     $<-3.10$   &     -    &   $-2.61$     &     0.04  &    $ -2.59$    &      0.11  &    0.90   &   0.21   \\
   2909-54653-0496  &     $<-3.20$   &     -    &   $-2.82$     &     0.03  &    $ -2.73$    &      0.13  &    0.30   &     -    \\
   2913-54526-0558  &     $ -3.04$   &  0.07    &   $-2.59$     &     0.01  &    $ -2.79$    &      0.09  &    0.03   &   0.68   \\
   2924-54582-0320  &     $<-3.10$   &     -    &   $-2.93$     &     0.07  &    $<-3.30$    &       -    &      -    &   1.21   \\
   2964-54632-0489  &     $ -3.14$   &  0.14    &   $-2.60$     &     0.01  &    $ -2.90$    &      0.18  &    0.50   &   0.17   \\
   2965-54594-0461  &     $ -3.12$   &  0.04    &   $-2.71$     &     0.02  &    $ -2.69$    &      0.09  &    0.35   &   0.91   \\
   3003-54845-0181  &     $ -3.27$   &  0.11    &   $-2.80$     &     0.01  &    $ -2.65$    &      0.10  &    0.49   &   0.95   \\
   3003-54845-0453  &     $<-3.70$   &     -    &   $-3.40$     &     0.02  &    $ -3.52$    &      0.17  &    0.06   &     -    \\
   3105-54825-0010  &     $ -3.21$   &  0.12    &   $-2.63$     &     0.03  &    $ -2.33$    &      0.17  &    0.25   &   0.56   \\
   3112-54802-0282  &     $ -3.01$   &  0.17    &   $-2.77$     &     0.04  &    $ -2.55$    &      0.16  &    0.51   &   0.25   \\
   3121-54749-0496  &     $ -3.27$   &  0.09    &   $-2.59$     &     0.01  &    $ -2.40$    &      0.11  &    0.14   &   0.61   \\
   3122-54821-0178  &     $<-3.28$   &     -    &   $-2.90$     &     0.02  &    $ -2.83$    &      0.13  &    0.49   &     -    \\
   3122-54821-0621  &     $ -3.12$   &  0.16    &   $-2.66$     &     0.02  &    $ -2.62$    &      0.15  &    0.59   &   0.37   \\
   3130-54740-0489  &     $ -3.31$   &  0.08    &   $-2.98$     &     0.01  &    $ -2.82$    &      0.11  &    0.37   &   0.96   \\
   3151-54804-0328  &     $ -3.15$   &  0.12    &   $-2.62$     &     0.03  &    $ -2.69$    &      0.17  &    0.26   &   0.42   \\
   3152-54801-0068  &     $ -3.34$   &  0.17    &   $-3.16$     &     0.04  &    $ -2.57$    &      0.17  &    0.30   &     -    \\
   3154-54821-0623  &     $<-3.20$   &     -    &   $-2.95$     &     0.05  &    $ -2.87$    &      0.21  &    0.31   &     -    \\
   3170-54859-0571  &     $ -3.21$   &  0.13    &   $-2.66$     &     0.02  &    $ -2.67$    &      0.13  &    0.61   &   1.12   \\
   3171-54862-0412  &     $ -3.26$   &  0.09    &   $-2.62$     &     0.01  &    $ -2.85$    &      0.17  &    0.24   &   0.67   \\
   3178-54848-0086  &     $<-3.40$   &     -    &   $-2.95$     &     0.03  &    $ -3.08$    &      0.09  &    0.37   &   0.60   \\
   3180-54864-0372  &     $<-3.30$   &     -    &   $-3.23$     &     0.02  &    $<-3.60$    &       -    &    0.90   &   2.57   \\
   3183-54833-0490  &     $<-3.40$   &     -    &   $-2.78$     &     0.02  &    $ -3.19$    &      0.13  &    0.69   &     -    \\
   3184-54850-0442  &     $<-3.27$   &     -    &   $-2.76$     &     0.04  &    $ -3.03$    &      0.11  &    0.40   &     -    \\
   3184-54850-0451  &     $ -3.11$   &  0.13    &   $-2.77$     &     0.02  &    $ -2.57$    &      0.03  &    0.26   &   0.24   \\
   3186-54833-0328  &     $<-3.30$   &     -    &   $-2.65$     &     0.04  &    $ -2.81$    &      0.25  &    0.49   &     -    \\
   3187-54821-0618  &     $ -3.08$   &  0.10    &   $-2.77$     &     0.02  &    $ -2.78$    &      0.09  &    0.18   &   1.15   \\
   3194-54833-0090  &     $<-3.00$   &     -    &   $-2.81$     &     0.04  &    $ -2.71$    &      0.23  &    0.57   &   2.74   \\
   3195-54832-0153  &     $<-3.50$   &     -    &   $-2.62$     &     0.04  &    $ -3.18$    &      0.23  &      -    &   1.23   \\
   3209-54906-0287  &     $ -3.26$   &  0.23    &   $-2.57$     &     0.04  &    $ -2.80$    &      0.24  &    0.27   &     -    \\
   3214-54866-0429  &     $<-3.30$   &     -    &   $-3.01$     &     0.02  &    $ -3.40$    &      0.29  &      -    &     -    \\
   3230-54860-0367  &     $<-3.40$   &     -    &   $-2.84$     &     0.02  &    $ -2.99$    &      0.12  &    0.02   &   0.78   \\
   3233-54891-0490  &     $ -3.53$   &  0.08    &   $-2.93$     &     0.01  &    $ -3.07$    &      0.10  &    0.31   &   0.95   \\
   3241-54884-0233  &     $ -3.08$   &  0.18    &   $-2.64$     &     0.03  &    $ -2.85$    &      0.15  &    0.49   &     -    \\
   3245-54894-0609  &     $ -3.15$   &  0.08    &   $-2.76$     &     0.02  &    $ -2.73$    &      0.13  &    0.33   &   0.65   \\
   3251-54882-0208  &     $<-3.21$   &     -    &   $-2.77$     &     0.03  &    $ -3.10$    &      0.19  &    0.31   &   0.82   \\
   3254-54889-0128  &     $<-3.00$   &     -    &   $-3.03$     &     0.05  &    $ -2.99$    &      0.20  &    0.90   &     -    \\
   3263-54887-0350  &     $ -3.23$   &  0.12    &   $-2.86$     &     0.03  &    $ -2.87$    &      0.23  &    0.87   &   1.00   \\
   3264-54889-0214  &     $ -3.28$   &  0.19    &   $-3.04$     &     0.06  &    $ -2.50$    &      0.22  &    0.25   &   0.44   \\
   3286-54910-0003  &     $<-3.00$   &     -    &   $-3.12$     &     0.12  &    $ -2.13$    &      0.28  &    0.53   &     -    \\
   3299-54908-0450  &     $ -3.07$   &  0.11    &   $-2.59$     &     0.02  &    $ -2.25$    &      0.17  &   $-0.02$ &  $-0.31$ \\
   3307-54970-0529  &     $ -3.39$   &  0.14    &   $-3.16$     &     0.01  &    $ -3.09$    &      0.12  &    0.43   &   1.16   \\
   3308-54919-0513  &     $ -3.01$   &  0.11    &   $-2.59$     &     0.02  &    $ -2.57$    &      0.18  &    0.24   &   0.05   \\
   3319-54915-0230  &     $ -3.53$   &  0.12    &   $-2.80$     &     0.03  &    $ -2.80$    &      0.16  &    0.41   &     -    \\
   3326-54943-0487  &     $<-3.09$   &     -    &   $-2.58$     &     0.02  &    $ -2.79$    &      0.17  &    0.04   &     -    \\
   3331-54977-0355  &     $<-3.00$   &     -    &   $-2.76$     &     0.04  &    $<-3.30$    &       -    &    0.22   &     -    \\
   3377-54950-0119  &     $<-3.10$   &     -    &   $-2.90$     &     0.08  &    $ -3.09$    &      0.22  &      -    &     -    \\
   3384-54948-0172  &     $ -3.25$   &  0.10    &   $-2.78$     &     0.02  &    $ -2.94$    &      0.17  &    0.17   &   0.94   \\
   3387-54951-0261  &     $<-3.20$   &     -    &   $-3.05$     &     0.09  &    $<-3.31$    &       -    &      -    &   1.39   \\
   3387-54951-0430  &     $ -3.14$   &  0.07    &   $-2.61$     &     0.02  &    $ -2.86$    &      0.14  &    0.44   &   0.58   \\
   3388-54947-0187  &     $<-3.10$   &     -    &   $-2.93$     &     0.06  &    $ -2.91$    &      0.19  &      -    &     -    \\
   3407-54971-0353  &     $ -3.03$   &  0.13    &   $-2.71$     &     0.04  &    $ -2.75$    &      0.20  &    0.06   &   1.10   \\
   3457-54984-0589  &     $ -3.07$   &  0.10    &   $-2.68$     &     0.01  &    $ -2.63$    &      0.09  &    0.36   &  $-0.03$ \\
   3639-55146-0544  &     $<-3.10$   &     -    &   $-2.86$     &     0.04  &    $ -3.16$    &      0.26  &    0.59   &     -    \\
   3647-55827-0970  &     $<-3.20$   &     -    &   $-2.58$     &     0.05  &    $ -3.06$    &      0.31  &   $-0.19$ &     -    \\
   3651-55247-0784  &     $<-3.25$   &     -    &   $-2.65$     &     0.04  &    $ -2.91$    &      0.22  &    0.00   &     -    \\
   3665-55247-0074  &     $ -3.31$   &  0.14    &   $-2.93$     &     0.01  &    $ -2.99$    &      0.14  &    0.02   &     -    \\
   4030-55634-0390  &     $<-3.10$   &     -    &   $-3.20$     &     0.05  &    $<-3.40$    &       -    &   $-0.50$ &     -    \\
   4041-55361-0401  &     $<-3.40$   &     -    &   $-3.42$     &     0.04  &    $ -3.37$    &      0.23  &    0.02   &     -    \\
   4043-55630-0998  &     $<-3.50$   &     -    &   $-3.93$     &     0.07  &    $<-3.70$    &       -    &      -    &   3.41   \\
   4084-55447-0163  &     $<-3.10$   &     -    &   $-3.05$     &     0.07  &    $<-3.40$    &       -    &      -    &     -    \\ 
   5176-56221-0514  &     $ -3.26$   &  0.25    &   $-2.82$     &     0.01  &    $ -3.35$    &      0.16  &   $-0.35$ &     -    \\
   5695-55978-0479  &     $<-3.10$   &     -    &   $-2.48$     &     0.16  &    $ -2.98$    &      0.29  &      -    &     -    \\
   5712-56602-0618  &     $ -3.27$   &  0.16    &   $-2.59$     &     0.03  &    $ -2.69$    &      0.17  &    0.34   &   0.56   \\
   5716-56684-0478  &     $<-3.30$   &     -    &   $-2.80$     &     0.05  &    $<-3.89$    &       -    &      -    &     -    \\
   5797-56273-0288  &     $<-3.20$   &     -    &   $-3.01$     &     0.04  &    $ -3.25$    &      0.29  &      -    &     -    \\
   6287-56221-0822  &     $<-3.00$   &     -    &   $-2.91$     &     0.07  &    $ -3.08$    &      0.36  &      -    &     -    \\
   6592-56535-0046  &     $<-3.00$   &     -    &   $-2.65$     &     0.04  &    $ -2.14$    &      0.40  &   $-0.20$ &     -    \\
   6604-56337-0611  &     $ -3.29$   &  0.12    &   $-2.76$     &     0.01  &    $ -2.89$    &      0.15  &    0.22   &     -    \\
\hline\hline
\end{longtable}
}

\section{Conclusions}
\label{conclusions}

We present stellar atmospheric parameters and abundance estimates of
[Fe/H], [Mg/H], and [Ca/H] for 108 extremely metal-poor stars from
SDSS/SEGUE and BOSS with iron abundances in the range $-4.0 <$ [Fe/H] $<
-3.0$. Below we summarize our main conclusions:

   \begin{enumerate}

\item The determination of Fe and Mg abundances for EMP stars
from individual lines in spectra at low spectral resolution ($R \sim
2000$) depends critically on the signal-to-noise ratio, due to the fact
that these lines are weak in this metallicity range. Conversely, the Ca
II HK resonance lines in the blue region of the spectrum are
sufficiently strong to reliably quantify the abundance of Ca. We
established a relation between S/N and [X/H] in order to determine upper
limits on the abundances of Fe, Ca, and Mg in cases where a reliable
estimate is not possible.

\item We derive an analytical function that reproduces the detection
limits for different elements as a function of S/N. These curves can be
well-fit by a parabola and a linear polynomial, with coefficients that
depend on $T_{\rm eff}$ and $\log g$. We report analytical functions
that specify the minimum S/N required to reliably estimate [Fe/H],
[Ca/H], and [Mg/H] in spectra with $R \sim 2000$ and S/N $<$ 90, from the spectral regions: 4885 $< \lambda <$ 5070 Å, 5220 $< \lambda <$ 5280 Å and 5295 $<
\lambda <$ 5500 Å (to determine Fe); 5160 $< \lambda <$ 5190 Å
(to determine Mg abundances); and 3910 $< \lambda <$ 3990 Å (for Ca abundances). 

\item We have determined [Ca/Fe] and [Mg/Fe] abundance ratios for our
program stars. The overall trend with metallicity is consistent with an
$\alpha$-element enhancement [$\alpha$/Fe]$\sim$+0.4~dex.
However, a number of stars for which only Fe upper limits were
estimated point to high [$\alpha$/Fe] ratios, mainly at the lowest
metallicities considered.

\item The [C/Fe] estimates obtained with the SSPP revealed a cumulative
frequency of $\sim26$\% CEMP stars for [Fe/H] $< -3.0$, comparable to that
found by Lee et al. (2013), 28\%. The frequency of CEMP stars also
increases with decreasing metallicity, as reported by previous studies.
We found no evidence for a [C/Fe] correlation with [Ca/Fe], [Mg/Fe], nor
the SSPP [$\alpha$/Fe] measurements. 

\item We have identified a bright ($g = 11.90$) EMP star in our
sample, SDSS J134144.61$+$474128.6, with enhanced [C/Fe] and [Mg/Fe], 
as well as a low absolute
carbon abundance, A(C) $= 6.11$, a pattern typically
associated with CEMP-no stars. Higher resolution spectroscopic follow-up
of this star is planned.

\end{enumerate}

There are a number of stars with reported [$\alpha$/Fe] significantly higher than $ +0.4$ in the
literature. For example, Aoki et al. (2007) reported on the highly
$\alpha$-element enhanced VMP star from the HK survey, BS~16934-002,
with [Mg/Fe] $= +1.23$, but ``normal'' [Ca/Fe] = +0.44, and no carbon
over-abundance. Other works reported high [C/Fe] and [Mg/Fe], but
little evidence of [Ca/Fe] enhancement with respect to the typical halo
values (Norris et al. 2013; Yong et al. 2013; Hansen et al. 2015. Two
scenarios were proposed for this chemical pattern: i) the results of
mixing and processing of material due to stellar rotation, or
ii) nucleosynthesis in mixing and fallback supernova models.
Elements such as Ca and Si provide the key to understanding which 
of these possibilities is more likely. We only found two stars for 
which both [C/Fe] and [Mg/Fe] are enhanced, but that exhibit normal
[Ca/Fe], the expected chemical pattern for a
massive spinstar (Norris et al. 2013; Maeder et al. 2015). 

Puzia et al. (2006) measured [$\alpha$/Fe] ratios significantly higher
than +0.5 for globular clusters in early-type elliptical galaxies based
on Lick line-index measurements, at $-1 < $[Z/H]\footnote{These authors
used [Z/H] to indicate the global metallicity in a galaxy, estimated
from Mg and Fe lines (see Puzia et al. 2006 and González 1993).}$<0$.
They suggest that massive stars are the potential progenitors, with M $>
20$M$_{\odot}$, or with M $\sim$130-190M$_{\odot}$ that explode as
pair-instability SNe. Both possibilities imply extremely short
timescales, on the order of few Myr. Therefore, they conclude that these
stars may belong to the first generation of star clusters formed in
their respective galaxies. In Paper~II we reported high [Ca/Fe] and
[Mg/Fe] median values for stars in the outer-halo region of the Galaxy,
at Galactocentric radii greater than 40 kpc. Such stars could have been
formed in low-mass fragments at very early stages of the evolution of
the Milky Way, and later accreted into the Galactic halo (see, e.g.,
Tissera et al. 2014, and references therein). 

In contrast, Caffau et al. (2013a) found three stars with [Fe/H] $< -3.0$
with low [$\alpha$/Fe] ratios. Similarly low [$\alpha$/Fe] stars had
been previously detected by Nissen and Schuster (2010). However, the
latter authors found these ratios for stars with higher [Fe/H], and
interpreted this population as being born after the explosion of Type Ia
SNe. In this scenario, the low [$\alpha$/Fe] ratios would then be the
result of the addition of Fe from low- to intermediate-mass stars.
However, at metallicities lower than [Fe/H] $= -3.0$, few SNIa
explosions are expected to have occurred. The interpretation offered by
Caffau et al. (2013) is that two starbursts could have taken place in
their progenitor fragment; the [low-$\alpha$/Fe] ratios could then have
resulted from gas enriched by SNIa explosions of stars formed in the
first burst. This hypothesis was also invoked by Carigi et al. (2002)
to explain low [O/Fe] in Milky Way dwarf spheroidal satellites from
chemical evolution models.

It would be desirable to obtain more accurate estimates to refine our
reported $\alpha$-element enhancements in cases where the iron abundance
could not be determined. The analysis performed in Paper~II revealed
high [$\alpha$/Fe] enhancements for very metal-poor stars in the
outer-halo region of our Galaxy. Accurate [Fe/H] estimates for our stars
would allow us to test whether these results are confirmed by our new
sample. 


\begin{acknowledgements}

E.F.A. acknowledges support from DGAPA-UNAM postdoctoral fellowships. C.A.P. acknowledges support from the Spanish
MINECO through grant AYA2014-56359-P. T.C.B. acknowledges partial support for this work from grants PHY
08-22648; Physics Frontier Center/Joint Institute or Nuclear
Astrophysics (JINA), and PHY 14-30152; Physics Frontier Center/JINA
Center for the Evolution of the Elements (JINA-CEE), awarded by the US
National Science Foundation. Y.S.L. acknowledges partial support from
the National Research Foundation of Korea to the Center for Galaxy
Evolution Research and Basic Science Research Program through the
National Research Foundation of Korea (NRF) funded by the Ministry of
Science, ICT \& Future Planning (NRF-2015R1C1A1A02036658).

Funding for SDSS-III has been provided by the Alfred P. Sloan
Foundation, the Participating Institutions, the National Science
Foundation, and the U.S. Department of Energy Office of Science. The
SDSS-III web site is http://www.sdss3.org/.

SDSS-III is managed by the Astrophysical Research Consortium for the
Participating Institutions of the SDSS-III Collaboration including the
University of Arizona, the Brazilian Participation Group, Brookhaven
National Laboratory, University of Cambridge, Carnegie Mellon
University, University of Florida, the French Participation Group, the
German Participation Group, Harvard University, the Instituto de
Astrofisica de Canarias, the Michigan State/Notre Dame/JINA
Participation Group, Johns Hopkins University, Lawrence Berkeley
National Laboratory, Max Planck Institute for Astrophysics, Max Planck
Institute for Extraterrestrial Physics, New Mexico State University, New
York University, Ohio State University, Pennsylvania State University,
University of Portsmouth, Princeton University, the Spanish
Participation Group, University of Tokyo, University of Utah, Vanderbilt
University, University of Virginia, University of Washington, and Yale
University. 

\end{acknowledgements}

\end{document}